\documentclass[10pt]{article} % For LaTeX2e
\usepackage[preprint]{tmlr}

\usepackage{amsmath,amsfonts,bm}

\def\eqref#1{equation~\ref{#1}}
\def\1{\bm{1}}

\DeclareMathAlphabet{\mathsfit}{\encodingdefault}{\sfdefault}{m}{sl}
\SetMathAlphabet{\mathsfit}{bold}{\encodingdefault}{\sfdefault}{bx}{n}

\usepackage{hyperref}
\usepackage{url}

\usepackage{amsmath}
\usepackage{amsthm} % theorem
\usepackage{bbold} % mathbb
\usepackage{booktabs}
\usepackage{xcolor}         % colors
\usepackage{graphicx}
\usepackage{subcaption}
\usepackage{stmaryrd} % for llbracket
\usepackage{multirow} % for multirow in tables
\usepackage{makecell}
\usepackage{pdflscape}
\usepackage{pbox}
\usepackage{color, colortbl} % rowcolor
\usepackage{algorithm}
\usepackage{algpseudocode}

\usepackage[capitalize,noabbrev]{cleveref}

\theoremstyle{plain}
\newtheorem{theorem}{Theorem}[section]

\theoremstyle{definition}
\newtheorem{definition}[theorem]{Definition}

\theoremstyle{remark}

\newtheorem{property}{Property}[section]

\definecolor{Gray}{gray}{0.9}

\title{Consistent Collaborative Filtering via Tensor Decomposition}

\author{\name Shiwen Zhao \email swzhao@apple.com \\
      \addr Apple
      \AND
      \name Charles G Crissman \email charleyc@gmail.com \\
      \addr Work done while at Apple
      \AND
      \name Guillermo R Sapiro \email gsapiro@apple.com\\
      \addr Apple}

\def\month{July}  % Insert correct month for camera-ready version
\def\year{2023} % Insert correct year for camera-ready version
\def\openreview{\url{https://openreview.net/forum?id=HqIuAzBxbh}} % Insert correct link to OpenReview for camera-ready version

\begin{document}

\maketitle

\begin{abstract}
  Collaborative filtering is the \emph{de facto} standard for analyzing users'
  activities and building recommendation systems for items. In this work we 
  develop {\it Sliced Anti-symmetric Decomposition} (SAD), a new model for 
  collaborative filtering based on implicit feedback. In contrast to traditional 
  techniques where a latent representation of users (user vectors) and items 
  (item vectors) are estimated, SAD introduces one additional latent vector to 
  each item, using a novel three-way tensor view of user-item interactions. 
  This new vector extends user-item preferences calculated by standard dot 
  products to general inner products, producing interactions between items when 
  evaluating their relative preferences. SAD reduces to state-of-the-art (SOTA) 
  collaborative filtering models when the vector collapses to $1$, 
  while in this paper we allow its value to be estimated from data. Allowing the
  values of the new item vector to be different from $1$ has profound implications.
  It suggests users may have \emph{nonlinear} mental models when evaluating items,
  allowing the existence of cycles in pairwise comparisons.   
  We demonstrate the efficiency of SAD in both simulated and real 
  world datasets containing over $1M$ user-item interactions. By comparing
  with seven SOTA collaborative filtering models with implicit feedbacks,
  SAD produces the most consistent personalized preferences, in the meanwhile
  maintaining top-level of accuracy in personalized recommendations. 
  We release the model and inference algorithms in a Python library \url{https://github.com/apple/ml-sad}.
\end{abstract}

\section{Introduction}
\label{sec:introduction}

Understanding preferences based on users' historical activities is key for 
personalized recommendation. This is particularly challenging when explicit 
ratings on many items are not available. In this scenario, historical activities 
are typically viewed as binary, representing whether a user has interacted with 
an item or not. Users' preferences must be inferred from such \emph{implicit} 
feedback with additional assumptions based on this binary data.

One common assumption is to view non-interacted items as negatives, meaning 
users are not interested in them; items that have been interacted are often 
assumed to be preferred ones \citep{hu2008collaborative, pan2008one}. In reality 
however, such an assumption is rarely met. For example, lack of interaction 
between a user and an item might simply be the result of lack of exposure. It is 
therefore more natural to assume that non-interacted items are a combination of 
the ones that users do not like and the ones that users are not aware of 
\citep{rendle_bpr_2009}. 

\begin{figure}[t]
\centering
\begin{subfigure}[b]{0.61\textwidth}
  \centering
  \includegraphics[width=\textwidth]{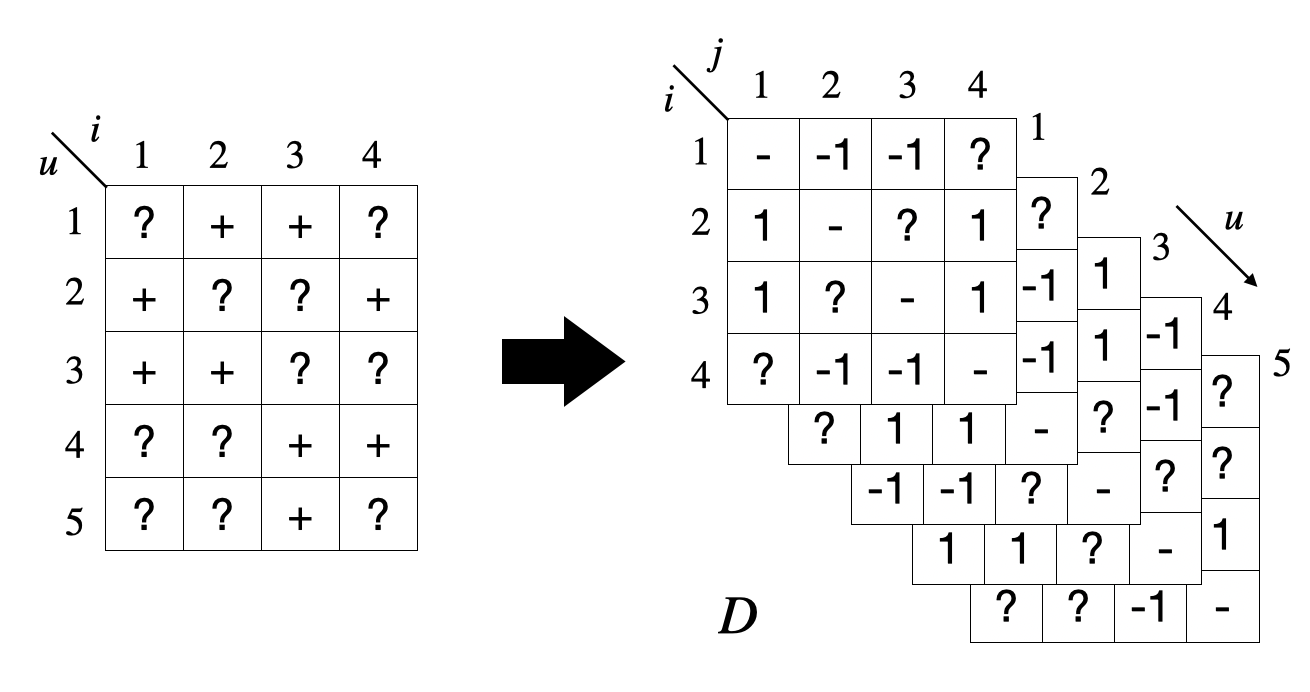}
  \caption{Data transformation to form three-way binary tensor $D$}
  \label{fig:tensor_D}
 \end{subfigure}
 \hfill
 \begin{subfigure}[b]{0.35\textwidth}
  \centering
  \includegraphics[width=\textwidth]{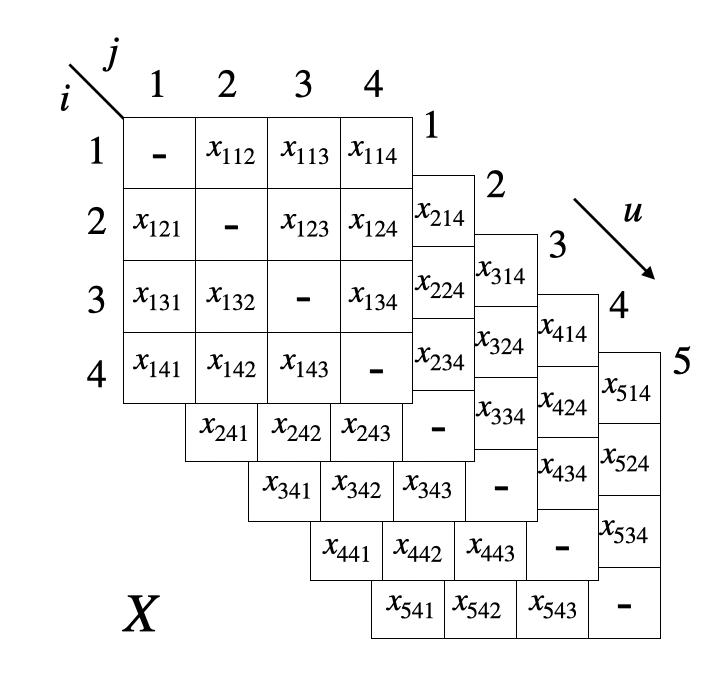}
  \caption{Tensor parameter $X$}
  \label{fig:tensor_X}
 \end{subfigure}
 \caption{Diagrams provide visualizations of both transformed observation $D$ 
 and parameters $X$ underlying the observation \citep{rendle_bpr_2009}.}
 \label{fig:tensor_viz}
\end{figure}

With this assumption, \citet{rendle_bpr_2009} proposed to give partial orders between items. 
Particularly, they assumed that items which users have interacted with are more 
preferable than the non-interacted ones. With this assumption in mind, Bayesian 
Personalized Ranking (BPR) was developed to perform personalized recommendations. 
In BPR, the observed data are transformed into a three-way binary tensor 
$D$ with the first dimension representing users. The other two dimensions represent 
items, encoding personalized preferences between item pairs (Figure \ref{fig:tensor_D}).
Mathematically, this means that any first order slice of $D$ at the $u$-th user, $D_{u::}$, is represented 
as a pairwise comparison matrix (PCM) between items. The $(i, j)$-th entry when observed, 
$d_{uij} \in \{-1, 1\}$, is binary, suggesting whether $u$-th user prefers ($d_{uij} = 1$)
the $i$-th item over the $j$-th one, or the other way around ($d_{uij} = -1$). 
The tensor $D$ is only partially observed, with 
missing entries where there is no prior knowledge to infer any preference for a 
particular user. The recommendation problem 
becomes finding a parsimonious parameterization of the generative model for observed 
entries in $D$ and estimating model parameters which best explain the observed data. 

The model used in BPR assumes that among the observed entries in $D$, the probability 
that the $u$-th user prefers the $i$-th item over the $j$-th item can be modeled as 
a Bernoulli distribution \citep{hu2008collaborative},
\begin{align} 
    \label{eq:bpr_p_uij}
    p(d_{uij} = 1 | X) = \frac{1}{1 + \exp{(-x_{uij} )}},
\end{align}
where $X = \{ x_{uij} \} \in \mathbb{R}^{n \times m \times m}$ is the collection of unknown
parameters of the Bernoulli distribution (Figure \ref{fig:tensor_X}). In fact, $x_{uij}$ is the natural parameter of a Bernoulli
distribution. \citet{rendle_bpr_2009} decompose the natural parameter by
\begin{align}
    \label{eq:bpr_x_uij}
    x_{uij} := x_{ui} - x_{uj} \in \mathbb{R}.
\end{align}

In the equation, $x_{ui}$ ($x_{uj}$) can be interpreted as the \emph{strength} of preference on the $i$-th ($j$-th) 
item for the $u$-th user. In other words, users' strengths of preference on different items are represented by 
scalar values denoted as $x_{ui}$. The relative preferences between items are therefore
characterized by the differences between their preference's strengths for a particular 
user. With this decomposition, $x_{uij}$ becomes $u$-th user's \emph{relative} preference on the $i$-th item over the 
$j$-th item. 

By further letting $x_{ui}$ be represented 
as a dot product between a $k$ dimensional user vector $\xi_u \in \mathbb{R}^k$ and an item vector $\eta_i \in \mathbb{R}^k$,
\begin{align}
    \label{eq:bpr_x_ui}
    x_{ui} = \langle \xi_u, \eta_i \rangle,
\end{align}
 \citet{rendle_bpr_2009} reveal the connection between their proposed BPR model and traditional collaborative filtering
 models such as matrix factorization.

The factorization in equations \ref{eq:bpr_x_uij} and \ref{eq:bpr_x_ui} provides a
parsimonious representation of the original Equation \ref{eq:bpr_p_uij}. Without the factorization,
one could model $x_{uij}$ independently. This model requires $n \times m \times m$ 
free parameters, and a model fitting would result in $x_{uij} = \infty$ for $d_{uij} = 1$, and
$x_{uij} = -\infty$ for $d_{uij} = -1$. The rests of $x_{uij}$ with missing $d_{uij}$ are left unknown.
The parsimonious representation, in contrast, reduces
the number of parameters to $(n + m) \times k$. The value of $x_{uij}$ now becomes coupled
with $x_{uit}$ for $t \ne j$, and entries in $D$ jointly impact parameter estimate of $x_{uij}$. 
This highlights a fundamental assumption behind collaborative filtering:
information can be learned from items that share similar pattern when they are interacted by users, 
and one user can learn from another user who interacts with similar set of items. 

Equation \ref{eq:bpr_x_ui} has a direct link to the Bradley-Terry model often studied 
when analyzing a PCM for decision making \citep{hunter2004mm, weng2011bayesian}. This 
model can be at least dated back to 1929
\citep{zermelo1929berechnung}. One property of the model is its transitivity: The 
relative preference $x_{uij}$ can be expressed as the sum of relative preferences of 
$x_{uit}$ and $x_{utj}$, for any user with $t \ne i$ and $t \ne j$. However, in 
reality this transitivity property is less frequently met. Only around $~3\%$ of real
world PCM's satisfy complete transitivity \citep{mazurek2017inconsistency}. The violation 
of the property is more conceivable when users are exposed to various types of items. 
For example, in an online streaming platform, one favorable movie could become less 
intriguing after a subscriber watches a different style/genre.

In this paper, we extend the original BPR model (which is one of the most fundamental 
models in recommendation systems) to allow non-transitive user ratings. In particular, we 
extend Equation \ref{eq:bpr_x_uij} to a more 
general form by proposing a new tensor decomposition. We denote our new model SAD 
(Sliced Anti-symmetric Decomposition). The new tensor decomposition introduces a second
set of non-negative item vectors $\tau_i$ for every item. Different from the first vector 
$\eta_i$, the new vector contributes negatively when calculating relative preferences,
producing counter-effects to the original strength of users' preferences; see Section 
\ref{sec:sad} for more details. Mathematically, the new vector extends the original
dot product in Equation \ref{eq:bpr_x_ui} to an \emph{inner product}. The original BPR 
model becomes a special case of SAD when the values in $\tau_i$ are all set to $1$, and 
the inner product reduces to a standard dot product. When $\tau_i$ contains entries that 
are not $1$, the transitivity property no longer necessarily holds. While assigning an 
$l_1$ regularization to the entries in $\tau_i$ to encourage its values being $1$ to 
reflect prior beliefs, SAD is able to infer its unknown value from real world data. 
We derive an efficient group coordinate descent algorithm for parameter estimation of
SAD. Our algorithm results in a simple stochastic gradient descent (SGD) producing fast and
accurate parameter estimations. Through a simulation study we first demonstrate the 
expressiveness of SAD and efficiency of the SGD algorithm. We then compare SAD to seven
alternative SOTA recommendation models in three publicly available datasets with distinct
characteristics. Results confirm that our new model permits to exploit information and 
relations between items not previously considered, and provides more consistent and accurate 
results as we will demonstrate in this paper. 

\section{Related Works}
\label{sec:relatedworks}

\textbf{Inferring priority via pairwise comparison.} The Bradley-Terry model 
\citep{hunter2004mm, weng2011bayesian, zermelo1929berechnung} has been heavily used along 
this line of research. In the Bradley-Terry model, the probability of the $i$-th unit 
(an individual, a team, or an item) being more preferable than the $j$-th unit (denoted 
as $i \succ j$) is modeled by 
\begin{align}
    \label{eq:bradley_terry}
    p(i \succ j) = \lambda_i/(\lambda_i + \lambda_j),
\end{align}
where $\lambda_i$ represents the strength, or degree, of preference of the $i$-th unit. 
The goal is to estimate $\lambda_i$ for all units based on pairwise comparisons. The 
link to Equation \ref{eq:bpr_p_uij} becomes clear once we omitting user index and 
set $x_i = \log(\lambda_i)$. In fact, the original BPR model can be viewed as an 
extension to the Bradley-Terry model to allow personalized parameters, and the strength 
of preferences are assumed to be dot products of user and item vectors as in Equation
\ref{eq:bpr_x_ui}.

Various algorithms have been developed for parameter estimation of this model. For 
example, \citet{hunter2004mm} developed a class of algorithms named 
minorization-maximization (MM) for parameter estimation. In MM, a minorizing function 
$Q$ is maximized to find the next parameter update at every iteration. Refer to the
work by \citet{hunter2004tutorial} for more details related to MM. \citet{weng2011bayesian}
proposed a Bayesian approximation method to estimate team's priorities from outputs of 
games between teams. Most recently, \citet{wang2021iterative} developed a bipartite graph 
iterative method to infer priorities from large and sparse pairwise comparison matrices. 
They applied the algorithm to the Movie-Lens dataset to rank movies based on their 
ratings aggregated from multiple users. Our paper is different from aforementioned 
models in that we model user-specific item preferences under personalized settings. 

\textbf{Tensor decompositions for recommendation.} Compared to traditional 
collaborative filtering methods using matrix factorizations, tensor decompositions have 
received less attention in this field until recently. The BPR model can be viewed as one of the first
attempts to approach the recommendation problems using tensor analysis. As discussed in 
Section \ref{sec:introduction}, by making the assumption that interacted items are more 
preferrable compared to non-interacted ones, user-item implicit feedback are 
represented as a three-way binary tensor \citep{rendle_bpr_2009}. In their later work, 
the authors developed tensor decomposition models for personalized tag recommendation 
\citep{rendle2010pairwise}. The relationship between their approach and traditional 
tensor decomposition approaches such as Tucker and PARAFAC (parallel factors) 
decompositions \citep{kiers2000towards, tucker1966some} was discussed. 

Recently, tensor decomposition methods have been used to build recommendation
systems using information from multiple sources. \citet{wermser2011modeling} 
developed a context aware tensor decomposition approach by using information from 
multiple sources, including time, location, and sequential information. 
\citet{hidasi2012fast} considered implicit feedback and incorporated contextual 
information using tensor decomposition. They developed an algorithm which scaled 
linearly with the number of non-zero entries in a tensor. A comprehensive review 
about applications of tensor methods in recommendation can be found by
\citet{frolov2017tensor} and references therein. Different from leveraging multiple
sources of information, the SAD model developed in this paper considers the basic 
scenario where only implicit feedback are available, the scenario that is considered 
in BPR model \citep{rendle_bpr_2009}. Our novelty lies in the fact that we propose a
more general form of tensor decomposition for modeling implicit feedback.

\textbf{Deep learning in recommendation models.} Deep learning has attracted 
significant attention in recent years, and the recommendation domain is no exception. 
Traditional approaches such as collaborative filtering and factorization machines 
(FM) have been extended to incorporate neural network components 
\citep{chen2017attentive, he2017neural, xiao2017attentional}. 
In particular, \citet{chen2017attentive} replaced the dot product that has been 
widely used in traditional collaborative filtering with a neural network containing 
Multilayer Perceptrons (MLP) and embedding layers. \citet{chen2017attentive} and
\citet{xiao2017attentional} introduced attention mechanisms \citep{vaswani2017attention} 
to both collaborative filtering and FM \citep{rendle2010factorization}.
Mostly recently \citet{rendle2020neural} revisited the comparison of traditional matrix 
factorization and neural collaborative filtering and concluded that matrix factorization
models can be as powerful as their neural counterparts with proper hyperparameters selected.
Despite the controversy, various types of deep learning models including convolutional 
networks, recurrent networks, variational auto-encoders (VAEs), attention models, and
combinations thereof have been successfully applied in recommendation systems. 
The work by \citet{zhang2019deep} provided an excellent review on this topic. This line of 
research doesn't have direct link to the SAD model considered in the current work. However,
we provide a brief review along this line since our model could be further extended to 
use the latest advances in the area.

\section{Notation}

We use $n$ to denote the total number of users in a dataset and $m$ to denote the total 
number of items. Users are indexed by $u \in [1, \cdots n]$. Items are indexed by both 
$i$ and $j \in [1, \cdots m]$. We use $k$ to denote the number of latent factors, and use 
$h \in [1, \cdots, k]$ to index a factor. Capital letters are used to denote a matrix or 
a tensor, and lowercase letters to denote a scalar or a vector. For example, the three-way 
tensor of observations is denoted as $D \in \mathbb{R}^{n\times m \times m}$ and the 
$(u, i, j)$-th entry is denoted as $d_{uij}$. Similarly, the user latent matrix is
denoted as $\Xi \in \mathbb{R}^{k \times n}$, and its $u$-th column is 
denoted as $\xi_u$ to represent the user vector for $u$-th user. We use $\xi^h$ to 
denote the $h$-th row (the $h$-th factor) of $\Xi$ viewed as a column vector.

\section{Tensor Sliced Anti-symmetric Decomposition}
\label{sec:sad}

We start with the original BPR model. The relative preference $x_{uij}$ defined in 
Equation \ref{eq:bpr_x_uij} forms a three-way tensor 
$X \in \mathbb{R}^{n\times m \times m}$. The BPR model \citep{rendle_bpr_2009} can be 
viewed as one parsimonious representation of tensor $X$. Let $\xi_u \in \mathbb{R}^k$ 
and $\eta_i \in \mathbb{R}^k$ denote the user and item vectors respectively, and let 
$\xi_{hu}$ ($\eta_{hi}$) indicate the $h$-th entry in $\xi_u$ ($\eta_i$). Equations
\ref{eq:bpr_p_uij}, \ref{eq:bpr_x_uij}, and \ref{eq:bpr_x_ui} can be re-written as
\begin{small}
\begin{equation*}
    \begin{aligned}
        X_{u::} = \sum_{h=1}^{k} \xi_{hu}(\widetilde{H}_h - (\widetilde{H}_h)^\top),
    \end{aligned}
\end{equation*}
\end{small}
where $X_{u::}$ is the first order slice of $X$ at $u$-th user, 
\begin{align*} 
    \widetilde{H}_h = \eta^h \circ \mathbb{1},
\end{align*}
$\eta^h \in \mathbb{R}^m$ being the $h$-th row of item matrix 
$H \in \mathbb{R}^{k\times m}$. $\mathbb{1} \in \mathbb{R}^m$ is used to denote a vector 
of all $1$'s and $\circ$ being the outer product. 

\subsection{Anti-symmetricity of $X_{u::}$}

As discussed in Section \ref{sec:relatedworks}, $X_{u::}$ represents a parsimonious
representation of parameters of a PCM. We formalize the property of $X$ as follows:
\begin{property}
    \label{antisymmetric_property}
    For every user $u$, the first order slice of $X$, is an anti-symmetric 
    with $X_{u::} = - X^\top_{u::}$.
\end{property}
This can be shown easily by letting $p_{uij} = p(d_{uij} = 1) = p(i \succ_u j)$ and 
noting that the relative preference $x_{uij}$ is the natural parameter of the 
corresponding Bernoulli distribution with $x_{uij} = \log(p_{uij}/(1 - p_{uij}))$.
Note that $x_{uij}$ is also known as the log-odds or logit. 

The decomposition introduced in BPR can be further written as
\small
\begin{align}\label{eq:bpr_tensor_slice}
    X_{u::} = \sum_{h=1}^k \xi_{hu} (\eta^h \circ \mathbb{1} - \mathbb{1} \circ \eta^h). 
\end{align}
\normalsize
Note that the anti-symmetricity is well respected in the equation. Intuitively, the
decomposition suggests that for the $u$-th user, her item preference matrix $X_{u::}$ 
can be decomposed as a weighted sum of $k$ anti-symmetric components, each of which 
is the difference of a rank one square matrix and its transpose.

\subsection{Generalization of BPR}

By replacing $\mathbb{1}$ with arbitrary vector $\tau^h \in \mathbb{R}^m$,
the new square matrix $\eta^h \circ \tau^h - \tau^h \circ \eta^h$ is still
anti-symmetric, and Property \ref{antisymmetric_property} still holds for the 
resulting $X_{u::}$. With this observation, we generalize Equation 
\ref{eq:bpr_tensor_slice} by proposing a new parsimonious representation of 
$X_{u::}$,
\small
\begin{align}\label{eq:sad}
    X_{u::} = \sum_{h=1}^{k} \xi_{hu} (\eta^h \circ \tau^h - \tau^h \circ \eta^h).
\end{align}
\normalsize
In this work we require entries in $\tau^h$ to be non-negative. The rationale will 
become clear in Section \ref{sec:sad_interpretation}. Furthermore, by letting 
$\Xi := (\xi_1, \xi_2, \cdots, \xi_n) \in \mathbb{R}^{k\times n}$, 
$H := (\eta^1, \eta^2, \cdots, \eta^k)^\top \in \mathbb{R}^{k\times m}$, and 
$T := (\tau^1, \tau^2, \cdots, \tau^k)^\top \in \mathbb{R}^{k\times m}_{+}$ (we use
$\mathbb{R}_{+}$ to denote the set of non-negative real numbers), we 
introduce the proposed Sliced Anti-symmetric Decomposition (SAD).

\begin{definition}
  We define the Sliced Anti-symmetric Decomposition (SAD) of $X$ to be the matrices
  $\Xi, H, T$ satisfying Equation \ref{eq:sad} above for every user index $u$.
  We denote this by
  \begin{align}   \label{eq:sad2}
      X \overset{\textrm{SAD}}{:=} \{ \Xi, H, T \}.
  \end{align}
\end{definition}

\subsection{Interpretation of SAD}
\label{sec:sad_interpretation}

To understand the interpretation of SAD, we start by re-writing
equations \ref{eq:bpr_x_uij} and \ref{eq:bpr_x_ui}  in BPR as
\small
\begin{align*}
    x_{uij} = \langle \xi_u, \eta_i \rangle - \langle \xi_u, \eta_j \rangle = 
    \sum_{h=1}^{k} (\xi_{hu} \eta_{hi} - \xi_{hu} \eta_{hj}).
\end{align*} 
\normalsize

The term $\xi_{hu} \eta_{hi}$ can be interpreted as the strength of preference 
of the $u$-th user on the $i$-th item from the $h$-th factor. The overall strength of 
preference, $x_{ui}$, is the sum of contributions from the $k$ individual factors. 
Accordingly, the relative preference over the $(i, j)$-th item pair for the $u$-th
user can be viewed as the difference between the preference strengths of the $i$-th
and the $j$-th items from the $u$-th user.

This interpretation has a direct link to the Bradley-Terry model (\ref{eq:bradley_terry}) 
as previously mentioned, in which 
the strength of the $i$-th item is described as a positive number $\lambda_i$.
Here the strength of preference is viewed as user specific and is represented by 
a real number $x_{ui} = \log{\lambda_{ui}}$.

SAD extends the original equations \ref{eq:bpr_x_uij} and \ref{eq:bpr_x_ui} 
by introducing a new non-negative vector $\tau_i$ for every item (a column in $T$
in Equation \ref{eq:sad2}). We can re-write 
Equation \ref{eq:sad} as follows for every item pair $i$ and $j$:
\begin{align}
    x_{uij} &= \langle \xi_u, \eta_i \rangle_{\textrm{diag}(\tau_j)} - \langle \xi_u, 
    \eta_j \rangle_{\textrm{diag}(\tau_i)} \nonumber \\
    & = \sum_{h=1}^{k} (\xi_{hu} \eta_{hi} \tau_{hj}
    - \xi_{hu} \eta_{hj} \tau_{hi}). \label{eq:sad_expand}
\end{align}

$\langle \cdot, \cdot \rangle_{\textrm{diag}(\tau_i)}$ in Equation \ref{eq:sad_expand}
denotes the inner product with a diagonal weight matrix having $\tau_i$ on the
diagonal. To be a proper inner product, we require $\tau_i$ to be
non-negative, resulting in a positive semi-definite matrix 
$\textrm{diag}(\tau_i)$.

The first term on the right hand side of Equation \ref{eq:sad_expand}, describing
the preference strength of the $i$-th item, now becomes dependent on $\tau_{hj}$, the 
$h$-th entry in $\tau_j$ of the $j$-th rival. When $\tau_{hj}$ is bigger than $1$, it 
increases the effect of $\xi_{hu}\eta_{hi}$. Similarly, the second term on the right 
hand side suggests that when $\tau_{hi}$ is bigger than $1$, it strengthens the effect 
of the $j$-th item. The opposites happen when either $\tau_{hj}$ or $\tau_{hi}$ is 
smaller than $1$. Therefore, while respecting the anti-symmetricity, the new non-negative
item vector $\tau_i$ can be viewed as a counter-effect acting upon the strength of relative
preferences, penalizing the strength when greater than $1$, while reinforcing when smaller than $1$.
In real world applications, a user's preference indeed may be influenced by different items. 
For example, during 
online shopping, one favorable dress may become less intriguing after a customer sees 
a different one with different style/color that matches her needs. In an online streaming
platform, one favorable movie could become less interesting after a subscriber watches 
another one with different style/genre. SAD allows us to capture these item-item interactions
by introducing a new set of vectors $\tau_i$. 

To summarize, we interpret the three factor matrices in SAD as follows:
\begin{itemize}
\item $\Xi$ represents the user matrix. Each user is represented by a user vector 
    $\xi_u \in \mathbb{R}^k$.
\item $H$ represents the \emph{left} item matrix, which is composed of \emph{left} item 
    vectors denoted as $\eta_i \in \mathbb{R}^k$. It contributes to the strength of 
    preference on the $i$-th item via an inner product with user vector $\xi_u$.
\item $T$ represents the \emph{right} item matrix, which contains non-negative \emph{right} 
    item vectors denoted as $\tau_i \in \mathbb{R}^k_{+}$. This set of vectors defines the 
    weight matrices of inner products between $\eta_i$ and $\xi_u$. It produces
    counter-effects to the original preference strengths, with values bigger than $1$ 
    adding additional strength to rival items in pairwise comparisons, and a value smaller 
    than $1$ producing the opposite effect. When $T=1$, the model reduces to the original 
    BPR model.
\end{itemize}

In SAD we estimate the value of $T$ from data. As discussed in Section \ref{sec:introduction},
we encourage the values of entries in $T$ to be $1$ unless there is strong evidence from the data
suggesting otherwise. This is achieved by adding an $l_1$ regularization centered around $1$ 
to the entries in $T$ independently. 

The $l_1$ regularization has another side effect. In Equation \ref{eq:sad_expand}, multiplying by
any constant $c$ to $\eta_{hi}$ and $1/c$ to $\tau_{hj}$ results in the same objective function, causing
$H$ and $T$ to be unidentifiable. The additional $l_1$ regularization around $1$ mitigates the identifiability 
problem by discouraging any constant multiplication that moves
$\tau_{hj}$ away from $1$, making the joint objective function identifiable between $H$ and $T$.

\subsection{The transitivity problem}
\label{subsec:transitivity}

In social science involving decision makings, PCMs have been investigated extensively
\citep{saaty2013analytic, wang2021iterative}. It is usually assumed that a PCM holds 
the transitivity property, resulting in the following observation introduced in Section
\ref{sec:introduction}: The relative preference of the $(i, j)$-th item pair can be 
derived from the sum of relative preferences of the $(i, t)$-th and $(t, j)$-th item 
pairs, with $t \ne i$ and $t \ne j$, $x_{uij} = x_{uit} + x_{utj}$. The original BPR 
meets this property nicely. After introducing $T$ in SAD, this property no longer
necessarily holds. One can show that $\tau_i = \tau_j = \tau_t$ for ternary $(i, j, t)$ 
is a sufficient condition for transitivity in SAD. In our model, we allow the violation of
this property, making the proposed model more realistic given the fact
that complete transitivity is met only in $3\%$ of real world applications 
\citep{mazurek2017inconsistency}.

\subsection{Inference algorithms}

To estimate model parameters, we maximize the log likelihood function directly. 
The log likelihood given observed entries in $D$ can be re-written as 
\small
\begin{align}\label{eq:sad_loglikelihood}
    \log & p(D | \Theta) = \sum_{(u, i, j)}\mathbf{1} (d_{uij}=-1)x_{uij} - \log{(1 + \exp{(-x_{uij})})},
\end{align}
\normalsize
where $\mathbf{1}(\cdot)$ is the indicator function, and the sum is taken with respect 
to non-missing entries in $D$ with $i < j$. Here we require $i < j$ to prevent us from 
double counting.

We take the derivatives with respect to the columns of $\Xi$, $H$, and $T$, resulting in
following gradients
\small
\begin{equation}\label{eq:sad_gradient}
    \begin{aligned}
        \frac{\partial \log p(D | \Theta)}{\partial \xi_u} &= w_{uij} (\eta_i \odot \tau_j - \eta_j \odot \tau_i), \\
        \frac{\partial \log p(D | \Theta)}{\partial \eta_i} &= w_{uij} \xi_{u} \odot \tau_j, \\
        \frac{\partial \log p(D | \Theta)}{\partial \eta_j} &= - w_{uij} \xi_{u} \odot \tau_i, \\
        \frac{\partial \log p(D | \Theta)}{\partial \tau_i} &= - w_{uij} \xi_{u} \odot \eta_j, \\
        \frac{\partial \log p(D | \Theta)}{\partial \tau_j} &= w_{uij} \xi_{u} \odot \eta_i,
    \end{aligned}
\end{equation}

\normalsize
from the $(u, i, j)$-th observation. 
Here $w_{uij} = \mathbf{1} (d_{uij}=-1) + \exp{(-x_{uij})}/(1 + \exp{(- x_{uij})})$
and $\odot$ is the element-wise product (Hadamard product).

Equation \ref{eq:sad_gradient} allows us to create a stochastic gradient descent (SGD) 
algorithm (Algorithm \ref{alg:sad}) to optimize the negative of the log likelihood. During optimization, 
we add an $l_1$ penalty with weight $w_1$ to the entries in $T$ independently to encourage 
their values to be $1$. In addition, 
we add an $l_2$ independent penalties with weight $w_2$ to both $\Xi$ and $H$ for 
further regularization. 

\begin{algorithm}
\caption{SGD for parameter estimation of SAD}\label{alg:sad}
\begin{algorithmic}
\Require $n$, $m$, $k$, $D \in \mathbb{R}^{n \times m \times m}$, $\rho$, $w_1$, $w_2$ \Comment{ $\rho$: \textrm{learning rate}, $w_1$, $w_2$ weights for $l_1$ and $l_2$}

\State \textrm{Initialization} $\Xi \in \mathbb{R}^{k \times n}, H \in \mathbb{R}^{k \times m}, T \in \mathbb{R}^{k \times m}_+$
\State $X = \{ \Xi, H, T \} $  \Comment{Equation \ref{eq:sad2}}
\While{\textrm{Convergence not met}}
\For{$u = 1 \cdots n$}
    \For{\textrm{Every item $i$ in interacted set}}
        \State \textrm{Random select item $j$ from non-interacted item set}
        \State \textrm{Calculate $\mathrm{d}\xi_{u}$, $\mathrm{d}\eta_{i}$, $\mathrm{d}\eta_{j}$, $\mathrm{d}\tau_{i}$, $\mathrm{d}\tau_{j}$} \Comment{Equation \ref{eq:sad_gradient}}
        \State $\xi_u \leftarrow \xi_u + \rho \cdot (\mathrm{d}\xi_{u} - 2 w_2 \xi_u$)
        \State $\eta_i \leftarrow \eta_i + \rho \cdot (\mathrm{d}\eta_{i} - 2 w_2 \eta_i$)
        \State $\eta_j \leftarrow \eta_j + \rho \cdot (\mathrm{d}\eta_{j} - 2 w_2 \eta_j$)
        \State $\tau_i \leftarrow \tau_i + \rho \cdot (\mathrm{d}\tau_{i} -  w_1 \mathbf{1}[\tau_i > 1] + w_1 \mathbf{1}[\tau_i < 1]$)   \Comment{$\mathbf{1}[\cdot]$: Entry-wise indicator to $\tau_i \in \mathbb{R}^k$}
        \State $\tau_j \leftarrow \tau_j + \rho \cdot (\mathrm{d}\tau_{j} -  w_1 \mathbf{1}[\tau_j > 1] + w_1 \mathbf{1}[\tau_j < 1]$) 
    \EndFor
\EndFor
\EndWhile
\end{algorithmic}
\end{algorithm}

We also develop an efficient Gibbs sampling algorithm for full posterior inference
under a Probit model setup. By drawing parameter samples from posterior distributions,
the Gibbs sampling algorithm has the advantage of producing accurate uncertainty estimation of 
the unknown parameters under Bayesian inference. We replace the logistic function in 
Equation \ref{eq:bpr_p_uij} with $p(d_{uij} = 1 | \Theta) = \Phi(x_{uij})$,
where $\Phi(x_{uij})$ is the cumulative distribution function (CDF) of the standard Guassian 
distribution centered at $x_{uij}$. By assigning spherical Gaussian priors to $\Xi$, $H$, 
and $T$, full conditional distributions can be derived. More details can be found
in Appendix.

\section{Simulation Study}
\label{simulation}

We first evaluate the performance of SAD and the SGD algorithm on simulation data, 
with the goal of examining the performance of our algorithm with true parameters known
ahead. We 
choose $n=20$ users, $m=50$ items, and $k=5$, resulting in $\Xi \in \mathbb{R}^{5\times 20}$, 
$H \in \mathbb{R}^{5 \times 50}$, and $T \in \mathbb{R}^{5\times 50}_{+}$. 
We consider two scenarios in the simulation. In the first 
simulation (Sim1) we set $T$ to $1$, effectively reducing SAD to the generative model of 
BPR \citep{rendle_bpr_2009}. In the second scenario (Sim2), we set a small proportion of 
$T$ to either $0.01$ or $5$, the other entries are set to $1$. For user matrix $\Xi$ and 
left item matrix $H$, their values are uniformly drawn from the interval $[-2, 2]$. We
calculate the preference tensor $X$ with Equation \ref{eq:sad_expand} and draw an
observation tensor $D$ from the corresponding Bernoulli distributions.

\begin{figure*}[th]
\centering
  \includegraphics[width=0.9\textwidth]{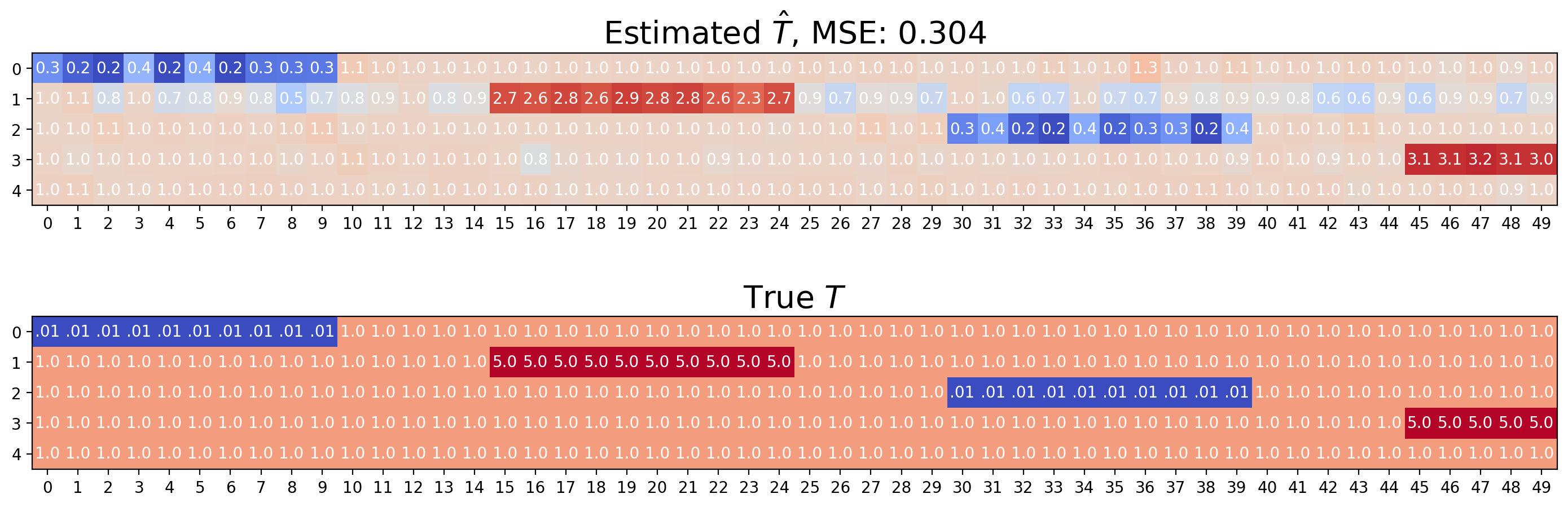}
  \caption{Comparison of $\hat{T}$ with ground truth. Factors are re-ordered in 
    $\hat{T}$ to match true $T$.}   
  \label{fig:comparision_T}
\end{figure*}

\begin{figure*}[th]
\centering
\includegraphics[width=0.9\textwidth]{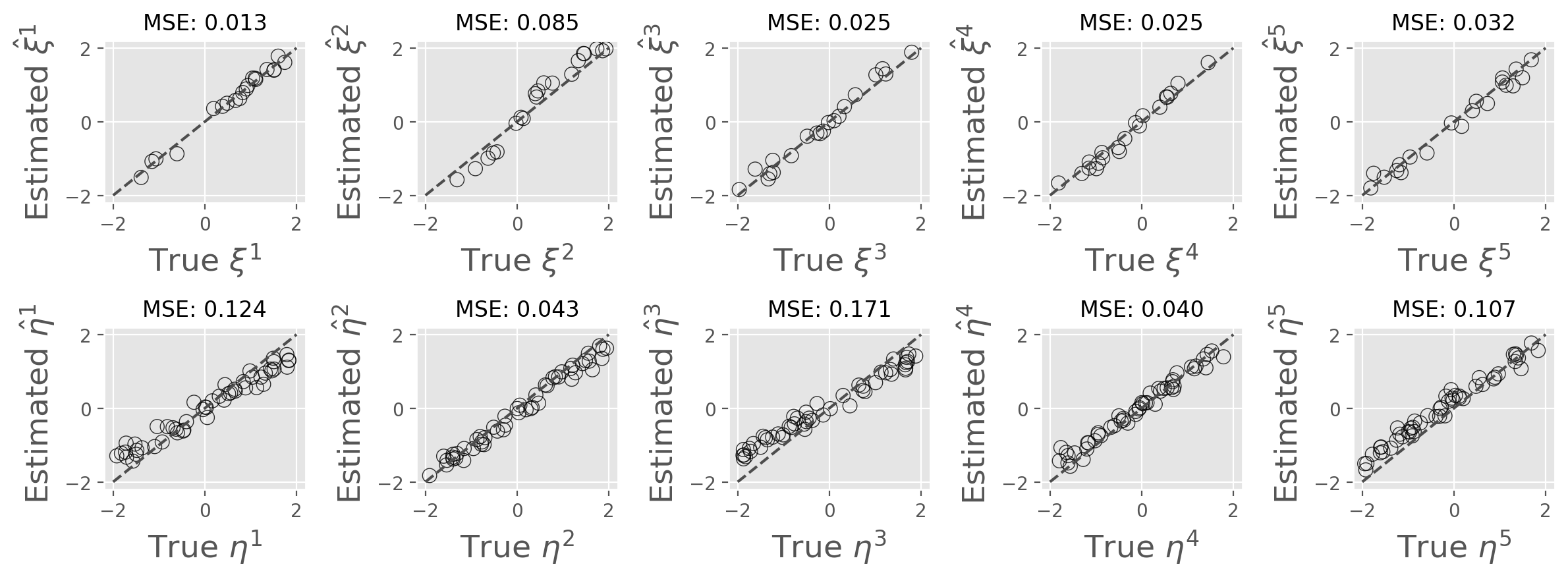}
\caption{Comparison of $\hat{\Xi}$, $\hat{H}$ with their ground truth. 
    Factors are subject to re-order and sign flips.}
\label{fig:comparison_XI_H}
\end{figure*}

We first examine the performance of SAD with complete observations in Sim2 to 
validate if our method is able to generate accurate parameter estimation. We run the
SGD algorithm with a learning rate $0.05$. The weight of the $l_1$ regularization 
assigned to $T$ is set to $0.01$, and the weight of the $l_2$ regularization is set to 
$0.005$. Initial values of parameters are randomly drawn from a standard Gaussian 
distribution. The number of latent factors $k$ is set to the true value. In reality when
$k$ is unknown, cross validation can be used to select the best value of $k$.
After 20 epochs, $\hat{T}$ is able to recover the sparse structure of the true parameters of $T$ up to
permutation of factors (Figure \ref{fig:comparision_T}). The user matrix and left item 
matrix converge to the true parameter values as well (Figure \ref{fig:comparison_XI_H}).

Next we examine the performance of SAD in both Sim1 and Sim2, under the scenarios with 
missing data. To be more specific, we randomly mark $x\%$ of $D$ as missing to mimic 
missing at random. Note that in the real world, observations could have more complex 
missingness structures. As a comparison, we run BPR under the same contexts.

\begin{figure}[t]
\centering
\begin{subfigure}[b]{0.48\textwidth}
  \centering
  \includegraphics[width=\textwidth]{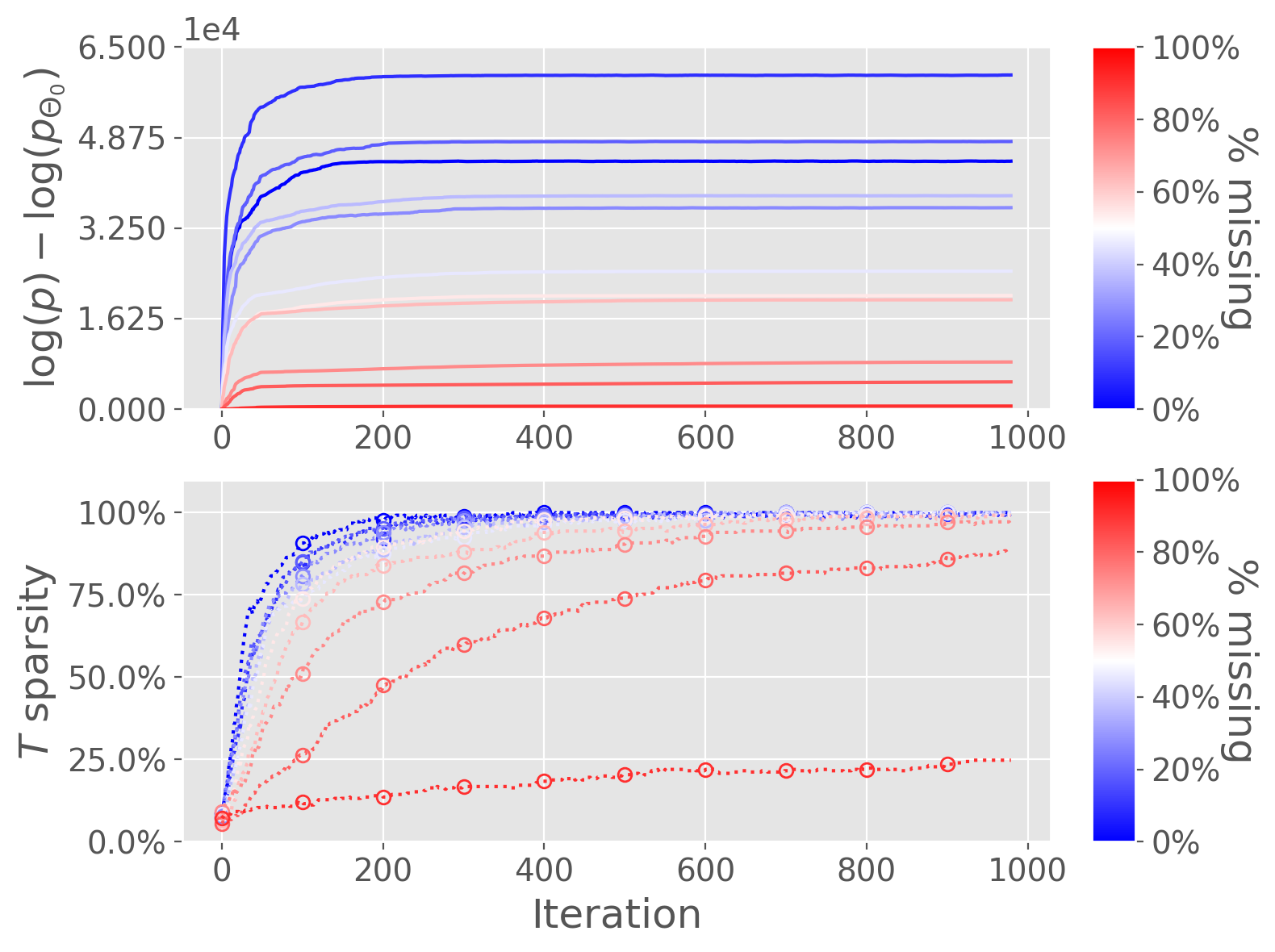}
  \caption{Likelilhood \& Sparsity (Sim1)}
  \label{fig:sad_missing_t_allone}
 \end{subfigure}
 \hfill
 \begin{subfigure}[b]{0.48\textwidth}
  \centering
  \includegraphics[width=\textwidth]{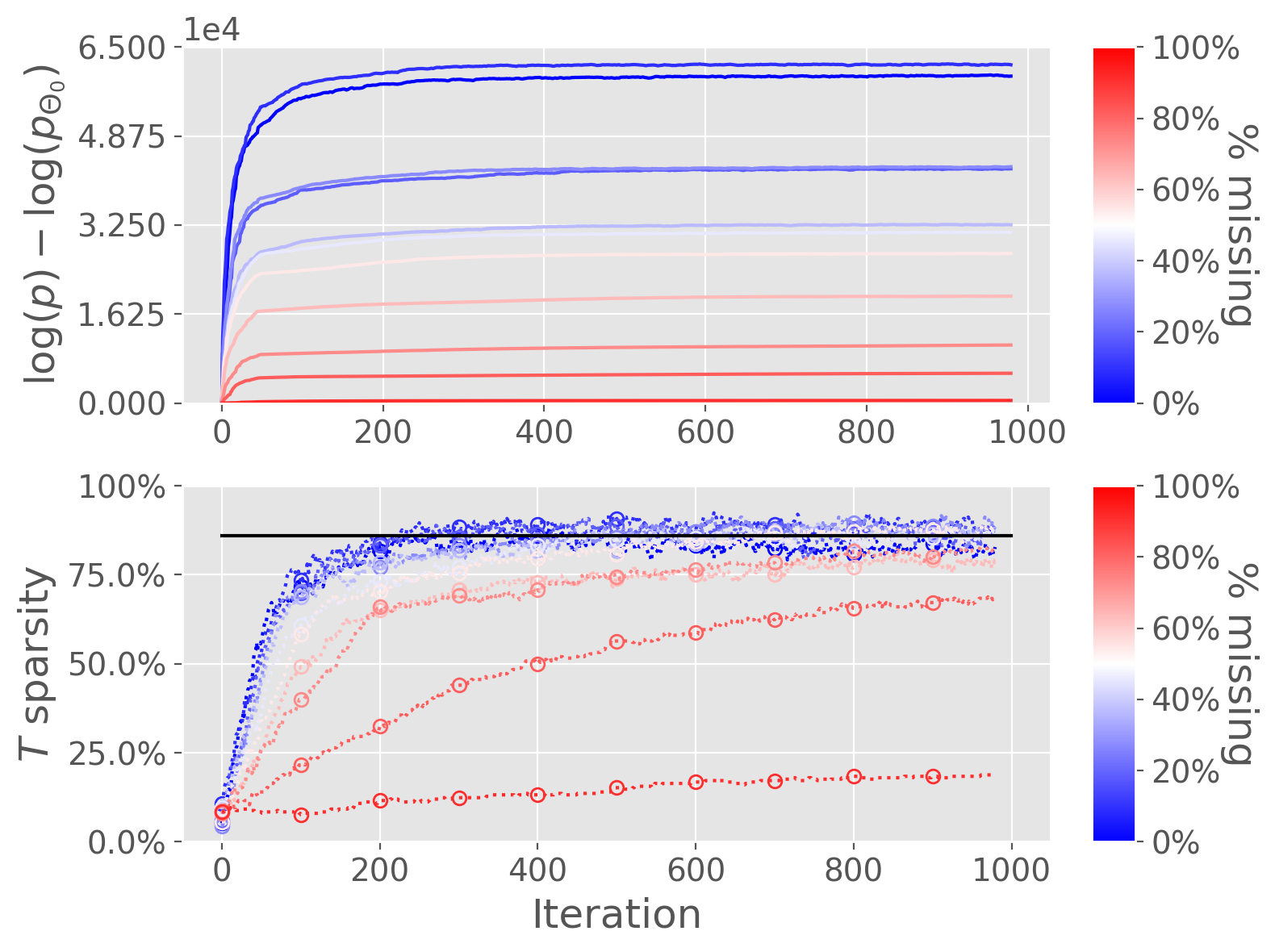}
  \caption{Likelilhood \& Sparsity (Sim2)}
  \label{fig:sad_missing_t_sparse}
 \end{subfigure}
 \hfill
 \begin{subfigure}[b]{0.48\textwidth}
  \centering
  \includegraphics[width=\textwidth]{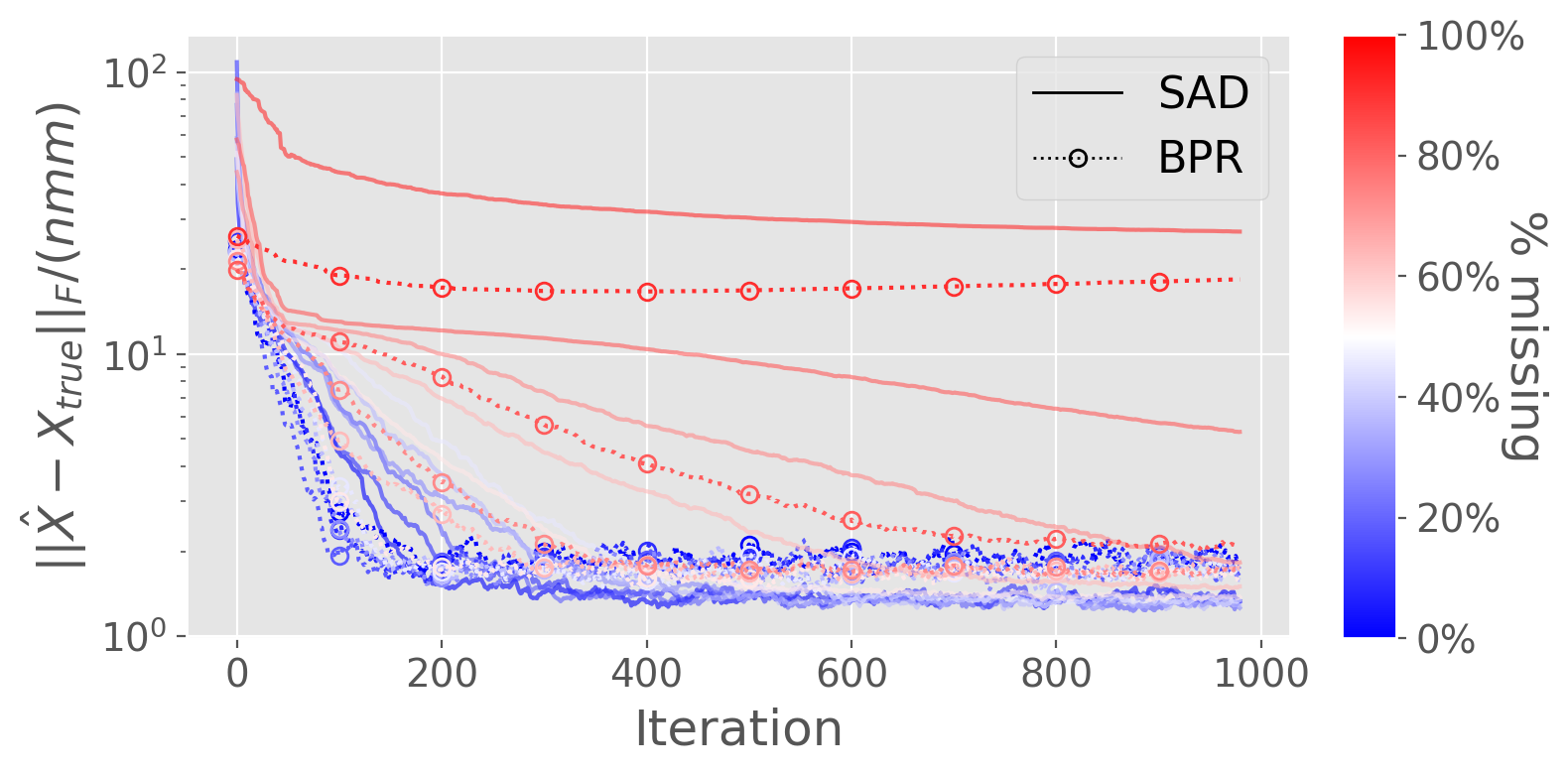}
  \caption{Convergence to True Parameter (Sim1)}
  \label{fig:sad_bpr_missing_mse_t_allone}
 \end{subfigure}
  \hfill
  \begin{subfigure}[b]{0.48\textwidth}
  \centering
  \includegraphics[width=\textwidth]{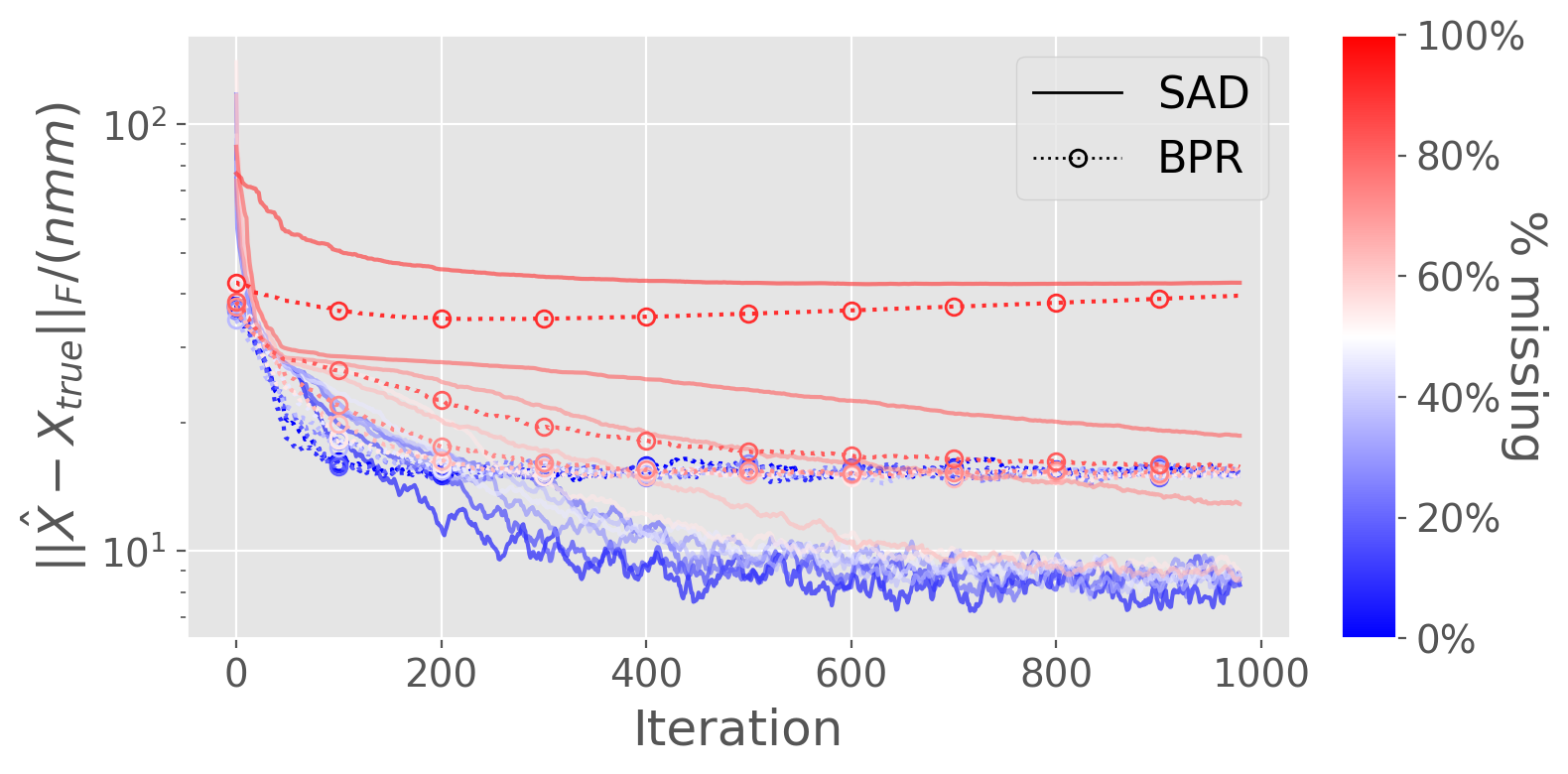}
  \caption{Convergence to True Parameter (Sim2)}
  \label{fig:sad_bpr_missing_mse_t_sparse}
 \end{subfigure}
 \caption{Convergence in Simulation Study. Top rows in \ref{fig:sad_missing_t_allone} and
   \ref{fig:sad_missing_t_sparse} show changes of log likelihood (Equation \ref{eq:sad_loglikelihood})
   during SGD optimization. $\log(p_{\Theta_0})$ is the log likelihood at $0$-th iteration.
   Bottom row of \ref{fig:sad_missing_t_sparse} shows the true
   sparsity ($86\%$) as black line.
   \ref{fig:sad_bpr_missing_mse_t_allone} and \ref{fig:sad_bpr_missing_mse_t_sparse} show
   the Frobenius distance between estimated parameter $\hat{X}$ and true parameter $X_{true}$
   during SGD optimization. Note that SAD (solid line) achieves a much lower distance compared 
   with BPR (dashed line) in \ref{fig:sad_bpr_missing_mse_t_sparse} under wide range of percentages of
   missingness.}
 \label{fig:sim-convergence}
\end{figure}

The convergences of SAD in Sim1 are shown in Figure \ref{fig:sad_missing_t_allone}, together 
with the estimated sparsity of $T$. Here the sparsity of $T$ is defined as the percentage of 
the entries in $T$ with $|\tau_{hj} - 1|<0.05$. When the percentage of missingness is 
at small or medium levels, SAD is able to converge to a sparsity close to $1$, suggesting the 
effectiveness of the $l_1$ regularization. It becomes more challenging when the percentage of 
missingness surpasses $70\%$. Figure \ref{fig:sad_bpr_missing_mse_t_allone} 
shows the trajectories of the mean squared error (MSE) between $\hat{X}$ and the true $X$ 
under different missingness percentages for both SAD and BPR. Both SAD and BPR are able to
converge to a low MSE with small/medium percentages of missingness. Similarly, with a high
percentage of missingness, the performance of both models begins to deteriorate. We conclude
that SAD has a performance \emph{on par} with BPR when data are simulated from the generative
model of BPR. 

Results for Sim2 are shown in Figure \ref{fig:sad_missing_t_sparse}. Note that the 
true sparsity of $T$ is $86\%$. SAD is able to generate an accurate estimation of the 
sparsity under small/medium percentages of missingness. When evaluating both models 
using MSE, SAD is able to achieve a much lower value due to its correct specification
of the generative model (Figure \ref{fig:sad_bpr_missing_mse_t_sparse}).

\section{Applications for Real Data}
\label{sec:application}

We select three real world datasets to evaluate SAD and compare against SOTA 
recommendation models. The datasets selected contain explicit integer valued ratings.
Nonetheless, we mask their values and view them as binary. The explicit ratings are
used as a means to evaluate models' consistency in pairwise comparison after model fitting
(details below). The first dataset used is from the Netflix Prize 
\citep{bennett2007netflix}. The original dataset contains movie ratings of $8,921$ 
movies from $478,533$ unique users, with a total number of ratings reaching
to over $50M$. We randomly select $10,000$ users as our first dataset. The resulting
dataset contains $8,693$ movies with over $1M$ ratings from the $10,000$ users. 
For the second dataset, we choose the Movie-Lens 1M dataset \citep{harper2015movielens}. 
It contains over $1M$ ratings from $6,040$ users on $3,706$ movies. 
As a third dataset, we consider the reviews of recipes from Food-Com 
\citep{majumder2019generating}. The complete dataset has $1.1M$ reviews from $227K$ 
users on $231K$ recipes. We select the top $20K$ users with the most activities, 
and filter out recipes receiving less than $50$ reviews. The resulting dataset
has $145K$ reviews from $17K$ users (users with zero activity are further removed 
after filtering recipes) on $1.4K$ recipes. 

The three datasets have distinct characteristics. Among the three datasets, the 
Food-Com review dataset has the least user-item interactions, even when the most 
active users/popular recipes are selected. The maximum number of items viewed by a 
single user is $878$. It also has the largest number of users. The Netflix dataset 
is the most skewed, with the number of items interacted by a user ranging from as 
low as $1$ to as high as over $8K$. The Movie-Lens dataset contains the largest 
number of user-item interactions, and is most uniformly distributed. Some details 
of the three datasets can be found in Table \ref{sec:appendix:datasets} in Appendix.

We choose seven SOTA recommendation models to compare with SAD. Their details are 
listed in Table \ref{table:alternatives}.
For each of the model considered, we perform a grid search to determine hyperparameters. 
Models are chosen based on their goodness-of-fit using log likelihood. We evaluate 
the models using a comprehensive leave one interaction out (LOO)
evaluation \citep{bayer2017generic, he2017neural, he2016fast}, in which we randomly
hold out one user-item interaction from training set for every user. Users who have 
only one interaction are skipped. We create $20$ such LOO sets for each dataset
considered. The dimension of latent space during evaluation is set to $500$ for all
models and datasets. The choice of the latent dimension can be further optimized using
methods such as across validation. In this work we choose the same number across
comparing SOTA models such that they can be evaluated on a common ground.

\begin{align}
\frac{1}{n \times (| I_u | - 1)} \sum_{u} \sum_{j \in I_u, j \ne o} & \big(\hat{x}_{hoj} \mathbf{1} (o \succ j) +  \hat{x}_{hjo}  \mathbf{1} (j \succ o) \big) \label{eq:eval_mean} \\
\frac{1}{n \times (| I_u | - 1)} \sum_{u} \sum_{j \in I_u, j \ne o} & \big(\mathbf{1} (\hat{x}_{hoj} > 0) \mathbf{1} (o \succ j) + \mathbf{1} (\hat{x}_{hjo} > 0)  \mathbf{1} (j \succ o) \big)  \label{eq:eval_percentage} 
\end{align}

We consider two aspects of model's performance during evaluation: consistency and 
recommendation. The consistency is defined as whether model's prediction matches 
user's actual pairwise preference. During evaluation, the hold out item $o$ and other 
items $j$ in interacted set $I_u$ of user $u$ are arranged such that $ o \succ_u j$  based
on users' actual ratings. The mean of their predicted preference (Equation \ref{eq:eval_mean}), 
the percentage of consistent predictions  (Equation \ref{eq:eval_percentage}), and the
median of per user percentage of consistency are reported in Table \ref{table:loo:evaluation}.
SAD has the most consistent results among all eight recommendation models except in 
one scenario, in which our model is second best.

\begin{table*}[th]
\caption{Model evaluations across $20$ LOO datasets. When evaluating consistency, item pairs between
hold out item $o$ and other interacted items $j$ for a user are ordered based on the user's actual ratings ($o \succ_u j$). 
The mean of their predicted preference (Equation  \ref{eq:eval_mean}), 
the percentage of predictions that match with actual ratings (Equation \ref{eq:eval_percentage}), and the median of per user 
percentage of match are reported. When evaluating a recommendation, the percentages of random hold out items 
that are ranked higher than $20$ (out of $100$) using two different ranking method (M1 and M2) are shown. 
See Table \ref{table:alternatives} in Appendix for details about each of the comparing models.}
\label{table:loo:evaluation}
\begin{center}\fontsize{9}{11}
\begin{small}
\begin{tabular}{llccccc}
\toprule
\multirow{2}{*}{Dataset} & \multirow{2}{*}{Model} & \multicolumn{3}{c}{Consistency} & \multicolumn{2}{c}{Recommendation} \\
\cmidrule(l{4pt}r{4pt}){3-5}
\cmidrule(l{4pt}r{4pt}){6-7}
 & & mean $x_{uij}$ & match ($\%$) & per user ($\%$) & M1 (\%) & M2 (\%) \\
\midrule
\multirow{8}{*}{Netflix} 
    & SAD & $\mathbf{0.024\pm 0.012}$ &  $\mathbf{33.7\pm 0.5}$ & $18.7\pm 0.3$ & $83.3 \pm 0.3$ & $83.9 \pm 0.3$ \\
    & BPR & $-0.019\pm 0.012$ & $32.6\pm 0.5$ & $12.7\pm 0.3$ & $81.8 \pm 0.4$ & $81.7 \pm 0.3$\\
    & SVD & $-0.824 \pm 0.032$ &  $10.1 \pm 0.1$ & $0.0 \pm 0.0$ & $2.0 \pm 0.5$ & $2.1 \pm 0.3$ \\
    & MF & $-0.018\pm 0.058$ & $8.1\pm0.5$ & $0.0\pm 0.0$ & $74.8 \pm 2.6$ & $45.4 \pm 8.7$ \\
    & PMF & $0.003\pm 0.015$ & $33.1\pm0.3$ & $\mathbf{26.9\pm 0.3}$ & $21.1 \pm 0.3$ & $20.2 \pm 0.3$\\
    & FM & $-0.538\pm 0.316$ & $12.9\pm0.2$ & $8.4\pm 0.1$ & $\mathbf{86.4 \pm 0.2}$ & $81.6 \pm 0.3$ \\
    & NCF & $-0.025\pm 0.043$ & $13.1\pm4.4$ & $5.6\pm 1.2$ & $86.0 \pm 0.3$ & $\mathbf{85.9 \pm 2.2}$ \\
    & $\beta$-VAE & $-0.011\pm 0.017$ & $21.1\pm0.9$ & $9.7\pm 2.1$ & $35.7 \pm 3.1$ & $31.0 \pm 5.7$  \\
\midrule
\multirow{8}{*}{\pbox{10cm}{Movie\\-Lens}} 
    & SAD & $\mathbf{0.120\pm0.014}$ & $\mathbf{36.4\pm0.6}$ & $\mathbf{29.9\pm 0.7}$ & $82.9 \pm 0.4$ & $82.1 \pm 0.4$ \\
    & BPR & $0.083 \pm 0.012$ & $35.7 \pm 0.6$ & $25.2 \pm 0.6$ & $77.7 \pm 0.5$ & $76.9 \pm 0.4$\\
    & SVD & $-0.324 \pm 0.011$ &  $7.0 \pm 0.4$ & $0.0 \pm 0.0$ & $3.5 \pm 0.1$ & $3.1 \pm 0.2$\\
    & MF & $0.024 \pm 0.103$ & $19.0 \pm 0.5$ & $0.0 \pm 0.0$ & $47.8 \pm 1.0$ & $27.0 \pm 0.7$\\
    & PMF & $0.027 \pm 0.016$ & $32.7 \pm 0.4$ & $26.4 \pm 0.4$ & $ 27.3 \pm 0.8$ & $22.8 \pm 0.5$\\
    & FM & $0.103 \pm 0.025$ & $21.7 \pm 0.3$ & $18.1 \pm 0.3$ & $78.8 \pm 0.3$ & $76.3 \pm 0.4$\\
    & NCF & $-0.241 \pm 0.346$ & $24.0 \pm 1.9$ & $14.5 \pm 2.0$ & $\mathbf{90.4 \pm 1.5}$ & $\mathbf{90.1 \pm 2.1}$\\
    & $\beta$-VAE & $-0.120 \pm 0.301$ & $13.2 \pm 2.1$ & $10.9 \pm 2.4$ & $71.0 \pm 0.3$ & $69.9 \pm 0.7$\\
\midrule
\multirow{8}{*}{\pbox{10cm}{Food\\-Com}}
    & SAD & $\mathbf{-0.329 \pm 0.002}$ & $\mathbf{14.9 \pm 0.5}$ & $0.0 \pm 0.0$ & $24.7 \pm 0.2$ & $23.9 \pm 0.2$\\
    & BPR & $-1.276 \pm 0.009$ & $5.9 \pm 0.5$ & $0.0 \pm 0.0$ & $23.2 \pm 0.3$ & $21.3 \pm 0.3$\\
    & SVD & $-5.152 \pm 0.039$ & $1.1 \pm 0.1$ & $0.0 \pm 0.0$ & $0.0 \pm 0.0$ & $0.0 \pm 0.0$ \\
    & MF & $ -1.956 \pm 0.161$ & $6.4 \pm 2.0$ & $0.0 \pm 0.0$ & $27.8 \pm 5.5$ & $24.6 \pm 6.1$\\
    & PMF & $ -0.434 \pm 0.031$ & $4.1 \pm 3.5$ & $0.0 \pm 0.0$ & $20.2 \pm 0.6$ & $21.0 \pm 1.1$\\
    & FM & $-5.324 \pm 0.247 $ & $9.3 \pm 1.7$ & $0.0 \pm 0.0$ & $35.9 \pm 0.2$ & $35.1 \pm 0.3$\\
    & NCF & $-11.362 \pm 0.852 $ & $8.9 \pm 0.7$ & $0.0 \pm 0.0$ & $\mathbf{38.2 \pm 4.8}$ & $\mathbf{37.1 \pm 5.5}$ \\
    & $\beta$-VAE & $-3.127 \pm 0.426$ & $10.2 \pm 1.1$ & $0.0 \pm 0.0$ & $16.3 \pm 3.1$ & $16.1 \pm 3.7$\\
\bottomrule
\end{tabular}
\end{small}
\end{center}
\end{table*}

To evaluate models' performance in recommendation, we create an item set containing $100$
non-interacted items by randomly sampling for each user. We combine the hold out item with
the $100$ items to form a test set, and examine whether models are able to rank the
hold out item high in the test set (Hit Ratio \citep{he2017neural}). SAD faces some unconventional challenges (hence opportunities)
in producing a ranking when violation of transitivity exists. Consider three items $i$, $j$ and $t$. With transitivity, if user 
has $i \succ j$ and $j \succ t$, then $i \succ t$ must hold. However, SAD can result in a scenario in which 
$t \succ i$, forming a preference cycle among the three items, in which case no ranking can be inferred.
We propose two methods using pairwise comparisons for SAD
in the evaluation. In the first method (M1), the number of non-interacted items in the test 
set that are more preferrable than the hold out item is dubbed as its rank. In a second 
method (M2), we use the ratings from interacted items kept in training set and calculate 
the proportion of interacted items that are less preferable than a test item. The 
proportion is used as a score to rank items in the test set. We calculate the 
percentage of hold out items that are ranked higher than $20$ in all the hold out 
items. SAD is among the top three best models that rank the hold out items in top
$20$ (Table \ref{table:loo:evaluation}). We argue that since our model's predictions 
match better with user's actual ratings, it is able to bring additional information by 
introducing diversity of items into recommendation, while respecting users's potential 
preferences. In Appendix, we illustrate examples in which SAD produces model 
predictions consistent with true ratings while other SOTA models fail.

\section{Discussions}

We proposed a new tensor decomposition model for collaborative 
filtering with implicit feedback. In contrast to traditional models, we introduced
a new set of non-negative latent vectors for items. While respecting
anti-symmetricity of parameters, the new vectors generalized the standard dot 
products for calculating user-item preferences to general inner products,
allowing items to interact when evaluating their relative preferences. When
such vectors were all set to $1$'s, our model reduced to standard 
collaborative filtering models. We allowed their values to be different
from $1$, enabling the violation of the known transitivity property.  The proposed
model generated accurate recommendations across multiple real world datasets
examined.

The existence of violation of transitivity has profound implication in real world applications.
The items that violate the property form cycles in the directed graph implied by pairwise
preferences of a user. For example, consider three items $i$, $j$ and $t$. Transitivity implies if
user $u$ has $i \succ j$ and $j \succ t$, then $t \succ i$ must hold. Violation of transitivity
suggests $t \succ i$, forming a preference cycle among the three items.
When SAD detects such violations, even a strong prior, $l_1$ 
regularization, suggesting otherwise, it indicates that users themselves may not have a
linear mental model in terms of ranking items. When making recommendations, the items
forming a cycle can be selected as a group, and users who are not certain which one in
the group is the most relevant can be exposed to the entire group, increasing the possibility
of producing a hit. 

When evaluating consistency in Section \ref{sec:application}, we used users' actual ratings
to determine pairwise preference between items. Many items in the datasets however had 
collided ratings due to the limited values those ratings can take. We chose the unevenly rated
items with partial orders as ground truth during evaluation. The partial orders are a proxy for users'
true preferences, and using them is the best we can do to evaluate whether our
model is able to produce consistent predictions. For evenly rated items, there is no ground truth
available to allow us to evaluate the performance. 

This topic touches one fundamental aspect in our existing rating system - using an integer
valued score with limited range as a sole reflection of user’s preference. Traditional collaborative
filtering models pander to this system, by assuming there is a linear ranking among items, and 
the ranking is dictated by a real number representing the preference strength. However, reality 
can hold evidence that violates the assumption. For example, \citet{mazurek2017inconsistency}
demonstrated only a small proportion of real world pairwise comparisons satisfy complete transitivity.
\citet{li_noisy_2013} showed the ubiquity of a user providing very different rating scores on closely
correlated items, producing self-contradictions. 
In addition to noisiness in user ratings, the observed self-contradictions is a manifestation of the collision
between user’s mental model and the rating system. SAD innovates
by providing a novel methodology to mathematically parameterize this mental model.

There can be parameterizations that both respect anti-symmetric property of $X_{u::}$, and
at the same time allow the violation of transitivity property other than
Equation \ref{eq:sad_expand}. For example, instead of introducing the new vector 
$\tau_i \in \mathbb{R}^k_+$ for every item, one can model $x_{uij}$ as

$$x_{uij} = \sum_{h=1}^k \xi_{uh} (\eta_{hi} - \eta_{hj}) \tau_{hij},$$

with $T_{h::} = \{\tau_{hij}\}_{1 \le i, j \le m} \in \mathbb{R}^{m \times m}$. $T_{h::} $ in this parameterization
can be viewed as an item interaction matrix for $h$-th factor. Additional structures can be assigned
to $T_{h::} $ to enable parsimoniousness, such as letting $T_{h::}$ be symmetric (in order to make $X_{u::}$ anti-symmetric)
and modeling it as a low rank matrix with $\tau_{hij} = \langle \alpha_{hi}, \beta_{hj} \rangle$, and $\alpha_{hi}, \beta_{hj} \in \mathbb{R}^\kappa$.
In our current work we focused on an efficient model with a simple 
interpretation. We delegate the research of exploring alternative parameterizations to future work.

In this paper, we restrict our scope to consider collaborative filterings for implicit feedbacks. SAD
can be applied to datasets beyond the scope. For instances, datasets with explicit ratings contain
partial orders that can be leveraged directly during model fitting, instead of being used to evaluate model
consistency in a post-hoc manner as in current work. Such datasets include the ones considered
in Section \ref{sec:application}, and numerous others such as 
\href{https://www.kaggle.com/competitions/yandex-personalized-web-search-challenge/overview}{Yandex challenge}
and \href{https://www.kaggle.com/c/kddcup2012-track2}{KDD Cup 2012}.
Other datasets that contain actual pairwise comparisons such as the one created by
\citet{pavlichenko_imdb-wiki-sbs_2021} are a natural fit to SAD as well. We expect the power of SAD
can be further enhanced with neural network components integrated.

\bibliography{reference}

\begin{thebibliography}{35}
\providecommand{\natexlab}[1]{#1}
\providecommand{\url}[1]{\texttt{#1}}
\expandafter\ifx\csname urlstyle\endcsname\relax
  \providecommand{\doi}[1]{doi: #1}\else
  \providecommand{\doi}{doi: \begingroup \urlstyle{rm}\Url}\fi

\bibitem[Bayer et~al.(2017)Bayer, He, Kanagal, and Rendle]{bayer2017generic}
Immanuel Bayer, Xiangnan He, Bhargav Kanagal, and Steffen Rendle.
\newblock A generic coordinate descent framework for learning from implicit
  feedback.
\newblock In \emph{Proceedings of the 26th International Conference on World
  Wide Web}, pp.\  1341--1350, 2017.

\bibitem[Bennett et~al.(2007)Bennett, Lanning, et~al.]{bennett2007netflix}
James Bennett, Stan Lanning, et~al.
\newblock The {Netflix} prize.
\newblock In \emph{Proceedings of {KDD} {C}up and {W}orkshop}, volume 2007,
  pp.\ ~35. New York, NY, USA., 2007.

\bibitem[Chen et~al.(2017)Chen, Zhang, He, Nie, Liu, and
  Chua]{chen2017attentive}
Jingyuan Chen, Hanwang Zhang, Xiangnan He, Liqiang Nie, Wei Liu, and Tat-Seng
  Chua.
\newblock Attentive collaborative filtering: {Multimedia} recommendation with
  item-and component-level attention.
\newblock In \emph{Proceedings of the 40th {International} {ACM} {SIGIR}
  {C}onference on {Research} and {Development} in {Information} {Retrieval}},
  pp.\  335--344, 2017.

\bibitem[Frolov \& Oseledets(2017)Frolov and Oseledets]{frolov2017tensor}
Evgeny Frolov and Ivan Oseledets.
\newblock Tensor methods and recommender systems.
\newblock \emph{Wiley {I}nterdisciplinary {R}eviews: {D}ata {M}ining and
  {K}nowledge {D}iscovery}, 7\penalty0 (3):\penalty0 e1201, 2017.

\bibitem[Graham et~al.(2019)Graham, Min, and Wu]{graham2019microsoft}
Scott Graham, Jun-Ki Min, and Tao Wu.
\newblock Microsoft recommenders: {T}ools to accelerate developing recommender
  systems.
\newblock In \emph{Proceedings of the 13th {ACM} {C}onference on {R}ecommender
  {S}ystems}, pp.\  542--543, 2019.

\bibitem[Harper \& Konstan(2015)Harper and Konstan]{harper2015movielens}
F~Maxwell Harper and Joseph~A Konstan.
\newblock The {M}ovielens datasets: {H}istory and context.
\newblock \emph{{ACM} {T}ransactions on {I}nteractive {I}ntelligent {S}ystems
  ({TIIS})}, 5\penalty0 (4):\penalty0 1--19, 2015.

\bibitem[He et~al.(2016)He, Zhang, Kan, and Chua]{he2016fast}
Xiangnan He, Hanwang Zhang, Min-Yen Kan, and Tat-Seng Chua.
\newblock Fast matrix factorization for online recommendation with implicit
  feedback.
\newblock In \emph{Proceedings of the 39th International ACM SIGIR conference
  on Research and Development in Information Retrieval}, pp.\  549--558, 2016.

\bibitem[He et~al.(2017)He, Liao, Zhang, Nie, Hu, and Chua]{he2017neural}
Xiangnan He, Lizi Liao, Hanwang Zhang, Liqiang Nie, Xia Hu, and Tat-Seng Chua.
\newblock Neural collaborative filtering.
\newblock In \emph{Proceedings of the 26th {I}nternational {C}onference on
  {W}orld {W}ide {W}eb}, pp.\  173--182, 2017.

\bibitem[Hidasi \& Tikk(2012)Hidasi and Tikk]{hidasi2012fast}
Bal{\'a}zs Hidasi and Domonkos Tikk.
\newblock Fast {ALS}-based tensor factorization for context-aware
  recommendation from implicit feedback.
\newblock In \emph{Joint {E}uropean {C}onference on {M}achine {L}earning and
  {K}nowledge {D}iscovery in {D}atabases}, pp.\  67--82. Springer, 2012.

\bibitem[Hu et~al.(2008)Hu, Koren, and Volinsky]{hu2008collaborative}
Yifan Hu, Yehuda Koren, and Chris Volinsky.
\newblock Collaborative filtering for implicit feedback datasets.
\newblock In \emph{2008 {E}ighth {IEEE} {I}nternational {C}onference on {D}ata
  {M}ining}, pp.\  263--272. Ieee, 2008.

\bibitem[Hug(2020)]{Hug2020}
Nicolas Hug.
\newblock Surprise: {A} {P}ython library for recommender systems.
\newblock \emph{Journal of {O}pen {S}ource {S}oftware}, 5\penalty0
  (52):\penalty0 2174, 2020.
\newblock \doi{10.21105/joss.02174}.
\newblock URL \url{https://doi.org/10.21105/joss.02174}.

\bibitem[Hunter(2004)]{hunter2004mm}
David~R Hunter.
\newblock {MM} algorithms for generalized {B}radley-{T}erry models.
\newblock \emph{The {A}nnals of {S}tatistics}, 32\penalty0 (1):\penalty0
  384--406, 2004.

\bibitem[Hunter \& Lange(2004)Hunter and Lange]{hunter2004tutorial}
David~R Hunter and Kenneth Lange.
\newblock A tutorial on {MM} algorithms.
\newblock \emph{The {A}merican {S}tatistician}, 58\penalty0 (1):\penalty0
  30--37, 2004.

\bibitem[Kiers(2000)]{kiers2000towards}
Henk~AL Kiers.
\newblock Towards a standardized notation and terminology in multiway analysis.
\newblock \emph{Journal of {C}hemometrics: {A} {J}ournal of the {C}hemometrics
  {S}ociety}, 14\penalty0 (3):\penalty0 105--122, 2000.

\bibitem[Li et~al.(2013)Li, Chen, Zhu, and Zhang]{li_noisy_2013}
Bin Li, Ling Chen, Xingquan Zhu, and Chengqi Zhang.
\newblock Noisy but non-malicious user detection in social recommender systems.
\newblock \emph{World Wide Web}, 16\penalty0 (5):\penalty0 677--699, November
  2013.
\newblock ISSN 1573-1413.
\newblock \doi{10.1007/s11280-012-0161-9}.
\newblock URL \url{https://doi.org/10.1007/s11280-012-0161-9}.

\bibitem[Liang et~al.(2018)Liang, Krishnan, Hoffman, and
  Jebara]{liang2018variational}
Dawen Liang, Rahul~G Krishnan, Matthew~D Hoffman, and Tony Jebara.
\newblock Variational autoencoders for collaborative filtering.
\newblock In \emph{Proceedings of the 2018 {W}orld {W}ide {W}eb {C}onference},
  pp.\  689--698, 2018.

\bibitem[Majumder et~al.(2019)Majumder, Li, Ni, and
  McAuley]{majumder2019generating}
Bodhisattwa~Prasad Majumder, Shuyang Li, Jianmo Ni, and Julian McAuley.
\newblock Generating personalized recipes from historical user preferences.
\newblock In \emph{Proceedings of the 2019 {C}onference on {E}mpirical
  {M}ethods in {N}atural {L}anguage {P}rocessing and the 9th {I}nternational
  {J}oint {C}onference on {N}atural {L}anguage {P}rocessing ({EMNLP-IJCNLP})},
  pp.\  5976--5982, 2019.

\bibitem[Mazurek \& Perzina(2017)Mazurek and Perzina]{mazurek2017inconsistency}
Ji{\v{r}}{\'\i} Mazurek and Radom{\'\i}r Perzina.
\newblock On the inconsistency of pairwise comparisons: {A}n experimental
  study.
\newblock \emph{Scientific {P}apers of the {U}niversity of {P}ardubice.
  {S}eries {D}, {F}aculty of {E}conomics and {A}dministration}, 2017.

\bibitem[Mnih \& Salakhutdinov(2008)Mnih and
  Salakhutdinov]{mnih2008probabilistic}
Andriy Mnih and Russ~R Salakhutdinov.
\newblock Probabilistic matrix factorization.
\newblock In \emph{Advances in {N}eural {I}nformation {P}rocessing {S}ystems},
  pp.\  1257--1264, 2008.

\bibitem[Pan et~al.(2008)Pan, Zhou, Cao, Liu, Lukose, Scholz, and
  Yang]{pan2008one}
Rong Pan, Yunhong Zhou, Bin Cao, Nathan~N Liu, Rajan Lukose, Martin Scholz, and
  Qiang Yang.
\newblock One-class collaborative filtering.
\newblock In \emph{2008 {E}ighth {IEEE} {I}nternational {C}onference on {D}ata
  {M}ining}, pp.\  502--511. IEEE, 2008.

\bibitem[Pavlichenko \& Ustalov(2021)Pavlichenko and
  Ustalov]{pavlichenko_imdb-wiki-sbs_2021}
Nikita Pavlichenko and Dmitry Ustalov.
\newblock {IMDB}-{WIKI}-{SbS}: An evaluation dataset for crowdsourced pairwise
  comparisons.
\newblock \emph{{arXiv} preprint {arXiv}:2110.14990}, 2021.

\bibitem[Rendle(2010)]{rendle2010factorization}
Steffen Rendle.
\newblock Factorization machines.
\newblock In \emph{2010 {IEEE} {International} {Conference} on {Data}
  {Mining}}, pp.\  995--1000, 2010.
\newblock \doi{10.1109/ICDM.2010.127}.

\bibitem[Rendle \& Schmidt-Thieme(2010)Rendle and
  Schmidt-Thieme]{rendle2010pairwise}
Steffen Rendle and Lars Schmidt-Thieme.
\newblock Pairwise interaction tensor factorization for personalized tag
  recommendation.
\newblock In \emph{Proceedings of the {T}hird {ACM} {I}nternational
  {C}onference on {Web} {S}earch and {D}ata {M}ining}, pp.\  81--90, 2010.

\bibitem[Rendle et~al.(2009)Rendle, Freudenthaler, Gantner, and
  Schmidt-Thieme]{rendle_bpr_2009}
Steffen Rendle, Christoph Freudenthaler, Zeno Gantner, and Lars Schmidt-Thieme.
\newblock {BPR}: {B}ayesian personalized ranking from implicit feedback.
\newblock In \emph{Proceedings of the {Twenty-Fifth} {Conference} on
  {Uncertainty} in {Artificial} {Intelligence}}, pp.\  452--461, 2009.

\bibitem[Rendle et~al.(2020)Rendle, Krichene, Zhang, and
  Anderson]{rendle2020neural}
Steffen Rendle, Walid Krichene, Li~Zhang, and John Anderson.
\newblock Neural collaborative filtering vs. matrix factorization revisited.
\newblock In \emph{Fourteenth {ACM} {C}onference on {R}ecommender {S}ystems},
  pp.\  240--248, 2020.

\bibitem[Saaty \& Vargas(2013)Saaty and Vargas]{saaty2013analytic}
Thomas~L Saaty and Luis~G Vargas.
\newblock The analytic network process.
\newblock In \emph{Decision {M}aking with the {A}nalytic {N}etwork {P}rocess},
  pp.\  1--40. Springer, 2013.

\bibitem[Salah et~al.(2020)Salah, Truong, and Lauw]{salah2020cornac}
Aghiles Salah, Quoc-Tuan Truong, and Hady~W Lauw.
\newblock Cornac: {A} comparative framework for multimodal recommender systems.
\newblock \emph{the Journal of {M}achine {L}earning {R}esearch}, 21:\penalty0
  1--5, 2020.

\bibitem[Tucker(1966)]{tucker1966some}
Ledyard~R Tucker.
\newblock Some mathematical notes on three-mode factor analysis.
\newblock \emph{Psychometrika}, 31\penalty0 (3):\penalty0 279--311, 1966.

\bibitem[Vaswani et~al.(2017)Vaswani, Shazeer, Parmar, Uszkoreit, Jones, Gomez,
  Kaiser, and Polosukhin]{vaswani2017attention}
Ashish Vaswani, Noam Shazeer, Niki Parmar, Jakob Uszkoreit, Llion Jones,
  Aidan~N Gomez, {\L}ukasz Kaiser, and Illia Polosukhin.
\newblock Attention is all you need.
\newblock In \emph{Advances in {N}eural {I}nformation {P}rocessing {S}ystems},
  pp.\  5998--6008, 2017.

\bibitem[Wang et~al.(2021)Wang, Kou, and Peng]{wang2021iterative}
Haomin Wang, Gang Kou, and Yi~Peng.
\newblock An iterative algorithm to derive priority from large-scale sparse
  pairwise comparison matrix.
\newblock \emph{{IEEE} {T}ransactions on {S}ystems, {M}an, and {C}ybernetics:
  {S}ystems}, 2021.

\bibitem[Weng \& Lin(2011)Weng and Lin]{weng2011bayesian}
Ruby~C Weng and Chih-Jen Lin.
\newblock A {B}ayesian approximation method for online ranking.
\newblock \emph{Journal of {M}achine {L}earning {R}esearch}, 12\penalty0 (1),
  2011.

\bibitem[Wermser et~al.(2011)Wermser, Rettinger, and
  Tresp]{wermser2011modeling}
Hendrik Wermser, Achim Rettinger, and Volker Tresp.
\newblock Modeling and learning context-aware recommendation scenarios using
  tensor decomposition.
\newblock In \emph{2011 {I}nternational {C}onference on {A}dvances in {S}ocial
  {N}etworks {A}nalysis and {M}ining}, pp.\  137--144. IEEE, 2011.

\bibitem[Xiao et~al.(2017)Xiao, Ye, He, Zhang, Wu, and
  Chua]{xiao2017attentional}
Jun Xiao, Hao Ye, Xiangnan He, Hanwang Zhang, Fei Wu, and Tat-Seng Chua.
\newblock Attentional factorization machines: {L}earning the weight of feature
  interactions via attention networks.
\newblock In \emph{Proceedings of the 26th {International} {Joint} {Conference}
  on {Artificial} {Intelligence}}, {IJCAI}'17, pp.\  3119--3125, Melbourne,
  Australia, 2017. AAAI Press.
\newblock ISBN 978-0-9992411-0-3.

\bibitem[Zermelo(1929)]{zermelo1929berechnung}
Ernst Zermelo.
\newblock Die berechnung der turnier-ergebnisse als ein maximumproblem der
  wahrscheinlichkeitsrechnung.
\newblock \emph{Mathematische {Z}eitschrift}, 29\penalty0 (1):\penalty0
  436--460, 1929.

\bibitem[Zhang et~al.(2019)Zhang, Yao, Sun, and Tay]{zhang2019deep}
Shuai Zhang, Lina Yao, Aixin Sun, and Yi~Tay.
\newblock Deep learning based recommender system: {A} survey and new
  perspectives.
\newblock \emph{{ACM} {C}omputing {S}urveys}, 52\penalty0 (1):\penalty0 1--38,
  2019.

\end{thebibliography}
\bibliographystyle{tmlr}

\newpage
\appendix

\section{Properties of Real World Datasets}
\label{sec:appendix:datasets}

\begin{table*}[!h]
\caption{Properties of the three real word datasets used in Section \ref{sec:application}}
\label{table:appendix:datasets}
\begin{center}
\begin{small}
\begin{tabular}{lccccc}
\toprule
\multirow{2}{*}{Dataset} & \multirow{2}{*}{\#Users}  & \multirow{2}{*}{\#Items} & \multirow{2}{*}{\#Ratings} & \multirow{2}{*}{Sparsity} & Quantiles of \#Ratings/User \\
                         &                          &                         &                           &               & $(\text{min}/5\%/50\%/95\%/\text{max})$ \\
\midrule
Netflix  & $10,000$ & $8,693$ &  $1,044,318$ & $98.80\%$ & $(1/6/46/400/8,237)$\\
Movie-Lens  & $6,040$ & $3,706$ &  $1,000,209$ & $95.53\%$ & $(20/23/96/556/2,314)$ \\
Food-Com & $17,482$ & $1,358$  & $145,431$ & $99.39\%$ & $(1/1/4/30/878)$ \\
\bottomrule
\end{tabular}
\end{small}
\end{center}
\end{table*}

\section{Gibbs Sampler for Posterior Inference of SAD}
\label{sec:appendix:gibbs}
We derive an efficient Gibbs sampling algorithm as a complement to the SGD algorithm 
in the main paper. The Gibbs sampling algorithm has the advantage of producing 
accurate uncertainty estimation of unknowns under Bayesian inference by
drawing parameter samples from the posterior distribution. The algorithm is an application 
of Bayesian Probit regression to the current setting. Specifically, we replace the original
logistic parameterization in Equation \ref{eq:bpr_p_uij} with the following equation:
\begin{equation}\label{eq:probit}
    p(d_{uij} = 1 | \Theta) := \Phi(x_{uij}),
\end{equation}
where $\Phi(x_{uij})$ is the CDF of a Gaussian distribution with mean
$x_{uij}$ and variance $1$. By augmenting the
model with a hidden tensor $Z = \{z_{uij}\}$, where 
$z_{uij} = x_{uij} + \epsilon_{uij}$ and 
$\epsilon_{uij} \overset{\textrm{i.i.d}}{\sim} N(0, 1)$, the Probit model is equivalent
to
\begin{equation*}
    d_{uij} = 
    \begin{cases}
    1 & z_{uij} > 0 \\
    -1 & z_{uij} \le 0
    \end{cases}.
\end{equation*}

With this new model, an efficient Gibbs sampling algorithm can be derived. As a toy 
example, we assign spherical Gaussian priors to rows of $\Xi$, $H$ and $T$ independently. 
With the likelihood defined in Equation \ref{eq:probit}, the following conditional 
posterior distributions can be derived.

\textbf{Posterior of $z_{uij}$}
\begin{equation*}
    z_{uij} | \Xi, H, T \sim 
    \begin{cases}
        N_{+}(x_{uij}, 1) & \textrm{if} ~d_{uij} = 1 \\
        N_{-}(x_{uij}, 1) & \textrm{if} ~d_{uij} = -1
    \end{cases},
\end{equation*}
where $N_{+}(\mu, \sigma)$ and  $N_{-}(\mu, \sigma)$ are truncated Gaussian distributions 
on positive and negative quadrants respectively. 

\textbf{Posterior of $\xi_u$ with $u = 1, \cdots, n$}
\begin{equation*}
     \xi_u | Z, \Xi \setminus \xi_u, H, T \sim N_{k}(\Sigma_u^\xi (\Psi_u^\xi)^\top \bar{z}_u^\xi, \Sigma_u^\xi),
\end{equation*}
where $(\Sigma_u^\xi)^{-1} = (\Psi_u^\xi)^\top (\Psi_u^\xi) + I $,
\begin{align*} 
    \Psi_u^\xi & = 
    \begin{pmatrix}
    \eta_{11}\tau_{12} - \eta_{12}\tau_{11} & \eta_{21}\tau_{22} - \eta_{22}\tau_{21} & \cdots & \eta_{k1}\tau_{k2} - \eta_{k2}\tau_{k1} \\
    \eta_{11}\tau_{13} - \eta_{13}\tau_{11} & \eta_{21}\tau_{23} - \eta_{23}\tau_{21} & \cdots & \eta_{k1}\tau_{k3} - \eta_{k3}\tau_{k1} \\
    \vdots & \vdots & \ddots & \vdots \\
    \eta_{12}\tau_{13} - \eta_{13}\tau_{12} & \eta_{22}\tau_{23} - \eta_{23}\tau_{22} & \cdots & \eta_{k2}\tau_{k3} - \eta_{k3}\tau_{k2} \\
    \vdots & \vdots & \ddots & \vdots \\
    \eta_{1,m-1}\tau_{1m} - \eta_{1m}\tau_{1,m-1} & \eta_{2,m-1}\tau_{2m} - \eta_{2m}\tau_{2,m-1} & \cdots & \eta_{k,m-1}\tau_{km} - \eta_{km}\tau_{k,m-1} \\
    \end{pmatrix} \\
    & \in \mathbb{R}^{m(m-1)/2 \times k},
\end{align*}
and $\bar{z}_u^\xi = [z_{u12}, z_{u13}, \cdots, z_{u23}, z_{u24}, \cdots, z_{u,m-1,m}]^\top \in \mathbb{R}^{m(m-1)/2}$.

\textbf{Posterior of $\eta_i$ with $i = 1, \cdots, m$}
\begin{equation*}
     \eta_i | Z, \Xi, H \setminus \eta_i, T \sim N_k(\Sigma_i^\eta (\Psi_i^\eta)^\top \bar{z}_i^\eta, \Sigma_i^\eta),
\end{equation*}
where $(\Sigma_i^\eta)^{-1} = (\Psi_i^\eta)^\top (\Psi_i^\eta) + I $, 

\begin{small}
\begin{align*} 
    \Psi_i^\eta & = 
    \begin{pmatrix}
    \xi_{11}\tau_{11}  & \xi_{21}\tau_{21} & \cdots & \xi_{k1}\tau_{k1} \\
    \xi_{11}\tau_{12}  & \xi_{21}\tau_{22} & \cdots & \xi_{k1}\tau_{k2}  \\
    \vdots & \vdots & \ddots & \vdots \\
    \xi_{11}\tau_{1,i-1}  & \xi_{21}\tau_{2,i-1} & \cdots & \xi_{k1}\tau_{k,i-1}  \\
    \xi_{11}\tau_{1,i+1}  & \xi_{21}\tau_{2,i+1} & \cdots & \xi_{k1}\tau_{k,i+1}  \\
     \vdots & \vdots & \ddots & \vdots \\
    \xi_{11}\tau_{1m}  & \xi_{21}\tau_{2m} & \cdots & \xi_{k1}\tau_{km}  \\
    \xi_{12}\tau_{11}  & \xi_{22}\tau_{21} & \cdots & \xi_{k2}\tau_{k1}  \\
    \vdots & \vdots & \ddots & \vdots \\
    \xi_{12}\tau_{1,i-1}  & \xi_{22}\tau_{2,i-1} & \cdots & \xi_{k2}\tau_{k,i-1}  \\
    \xi_{12}\tau_{1,i+1}  & \xi_{22}\tau_{2,i+1} & \cdots & \xi_{k2}\tau_{k,i+1}  \\
    \vdots & \vdots & \ddots & \vdots \\
    \xi_{12}\tau_{1m}  & \xi_{22}\tau_{2m} & \cdots & \xi_{k2}\tau_{km}  \\
    \vdots & \vdots & \ddots & \vdots \\
    \xi_{1n}\tau_{11}  & \xi_{2n}\tau_{21} & \cdots & \xi_{kn}\tau_{k1}  \\
    \vdots & \vdots & \ddots & \vdots \\
    \xi_{1n}\tau_{1,i-1}  & \xi_{2n}\tau_{2,i-1} & \cdots & \xi_{kn}\tau_{k,i-1}  \\
    \xi_{1n}\tau_{1,i+1}  & \xi_{2n}\tau_{2,i+1} & \cdots & \xi_{kn}\tau_{k,i+1}  \\
    \vdots & \vdots & \ddots & \vdots \\
    \xi_{1n}\tau_{1m}  & \xi_{2n}\tau_{2m} & \cdots & \xi_{kn}\tau_{km}  \\
    \end{pmatrix} \in \mathbb{R}^{n(m-1) \times k},
\end{align*}
\end{small}

\begin{small}
\begin{align*} 
    \bar{z}_i^\eta & = 
    \begin{pmatrix}
    z_{1i1} + \sum_{h=1}^k \xi_{h1} \tau_{hi} \eta_{h1}  \\
    z_{1i2} + \sum_{h=1}^k \xi_{h1} \tau_{hi} \eta_{h2}  \\
    \vdots \\
    z_{1,i,i-1} + \sum_{h=1}^k \xi_{h1} \tau_{hi} \eta_{h,i-1}  \\
    z_{1,i,i+1} + \sum_{h=1}^k \xi_{h1} \tau_{hi} \eta_{h,i+1}  \\
    \vdots \\
    z_{1im} + \sum_{h=1}^k \xi_{h1} \tau_{hi} \eta_{h,m}  \\
    z_{2i1} + \sum_{h=1}^k \xi_{h2} \tau_{hi} \eta_{h1}  \\
    \vdots \\
    z_{2,i,i-1} + \sum_{h=1}^k \xi_{h2} \tau_{hi} \eta_{h,i-1}  \\
    z_{2,i,i+1} + \sum_{h=1}^k \xi_{h2} \tau_{hi} \eta_{h,i+1}  \\
    \vdots \\
    z_{2im} + \sum_{h=1}^k \xi_{h2} \tau_{hi} \eta_{hm}  \\
    \vdots \\
    z_{ni1} + \sum_{h=1}^k \xi_{hn} \tau_{hi} \eta_{h1}  \\
    \vdots \\
    z_{n,i,i-1} + \sum_{h=1}^k \xi_{hn} \tau_{hi} \eta_{h,i-1}  \\
    z_{n,i,i+1} + \sum_{h=1}^k \xi_{hn} \tau_{hi} \eta_{h,i+1}  \\
    \vdots \\
    z_{nim} + \sum_{h=1}^k \xi_{hn} \tau_{hi} \eta_{hm}  \\
    \end{pmatrix} \in \mathbb{R}^{n(m-1)}.
\end{align*}
\end{small}

In the above notation, we take advantage of the anti-symmetric property of 
$Z_{u::}$ and assume $z_{uij} = - z_{uji}$ when $i > j$.

\textbf{Posterior of $\tau_j$ with $j = 1, \cdots, m$}
\begin{equation*}
     \tau_j | Z, \Xi, H, T \setminus \tau_j \sim N^{+}_k(\Sigma_j^\tau (\Psi_j^\tau)^\top \bar{z}_j^\tau, \Sigma_j^\tau),
\end{equation*}
where $(\Sigma_j^\tau)^{-1} = (\Psi_j^\tau)^\top (\Psi_j^\tau) + I $,
\begin{small}
\begin{align*} 
    \Psi_j^\tau & = 
    \begin{pmatrix}
    \xi_{11}\eta_{11}  & \xi_{21}\eta_{21} & \cdots & \xi_{k1}\eta_{k1} \\
    \xi_{11}\eta_{12}  & \xi_{21}\eta_{22} & \cdots & \xi_{k1}\eta_{k2}  \\
    \vdots & \vdots & \ddots & \vdots \\
    \xi_{11}\eta_{1,j-1}  & \xi_{21}\eta_{2,j-1} & \cdots & \xi_{k1}\eta_{k,j-1}  \\
    \xi_{11}\eta_{1,j+1}  & \xi_{21}\eta_{2,j+1} & \cdots & \xi_{k1}\eta_{k,j+1}  \\
    \vdots & \vdots & \ddots & \vdots \\
    \xi_{11}\eta_{1m}  & \xi_{21}\eta_{2m} & \cdots & \xi_{k1}\eta_{km}  \\
    \xi_{12}\eta_{11}  & \xi_{22}\eta_{21} & \cdots & \xi_{k2}\eta_{k1}  \\
    \vdots & \vdots & \ddots & \vdots \\
    \xi_{12}\eta_{1,j-1}  & \xi_{22}\eta_{2,j-1} & \cdots & \xi_{k2}\eta_{k,j-1}  \\
    \xi_{12}\eta_{1,j+1}  & \xi_{22}\eta_{2,j+1} & \cdots & \xi_{k2}\eta_{k,j+1}  \\
    \vdots & \vdots & \ddots & \vdots \\
    \xi_{12}\eta_{1m}  & \xi_{22}\eta_{2m} & \cdots & \xi_{k2}\eta_{km}  \\
    \vdots & \vdots & \ddots & \vdots \\
    \xi_{1n}\eta_{11}  & \xi_{2n}\eta_{21} & \cdots & \xi_{kn}\eta_{k1}  \\
    \vdots & \vdots & \ddots & \vdots \\
    \xi_{1n}\eta_{1,j-1}  & \xi_{2n}\eta_{2,j-1} & \cdots & \xi_{kn}\eta_{k,j-1}  \\
    \xi_{1n}\eta_{1,j+1}  & \xi_{2n}\eta_{2,j+1} & \cdots & \xi_{kn}\eta_{k,j+1}  \\
    \vdots & \vdots & \ddots & \vdots \\
    \xi_{1n}\eta_{1m}  & \xi_{2n}\eta_{2m} & \cdots & \xi_{kn}\eta_{km}  \\
    \end{pmatrix} \in \mathbb{R}^{n(m-1) \times k},
\end{align*}
\end{small}

\begin{small}
\begin{align*} 
    \bar{z}_j^\tau & = 
    \begin{pmatrix}
    - z_{1j1} + \sum_{h=1}^k \xi_{h1} \tau_{h1} \eta_{hj}  \\
    - z_{1j2} + \sum_{h=1}^k \xi_{h1} \tau_{h2} \eta_{hj}  \\
    \vdots \\
    - z_{1j,j-1} + \sum_{h=1}^k \xi_{h1} \tau_{h,j-1} \eta_{hj}  \\
    - z_{1j,j+1} + \sum_{h=1}^k \xi_{h1} \tau_{h,j+1} \eta_{hj}  \\
    \vdots \\
    - z_{1jm} + \sum_{h=1}^k \xi_{h1} \tau_{hm} \eta_{hj}  \\
    - z_{2j1} + \sum_{h=1}^k \xi_{h2} \tau_{h1} \eta_{hj}  \\
    \vdots \\
    - z_{2j,j-1} + \sum_{h=1}^k \xi_{h2} \tau_{h,j-1} \eta_{hj}  \\
    - z_{2j,j+1} + \sum_{h=1}^k \xi_{h2} \tau_{h,j+1} \eta_{hj}  \\
    \vdots \\
    - z_{2jm} + \sum_{h=1}^k \xi_{h2} \tau_{hm} \eta_{hj}  \\
    \vdots \\
    - z_{n,j,j-1} + \sum_{h=1}^k \xi_{hn} \tau_{h,j-1} \eta_{hj}  \\
    - z_{n,j,j+1} + \sum_{h=1}^k \xi_{hn} \tau_{h,j+1} \eta_{hj}  \\
    \vdots \\
    - z_{njm} + \sum_{h=1}^k \xi_{hn} \tau_{hm} \eta_{hj}  \\
    \end{pmatrix} \in \mathbb{R}^{n(m-1)}.
\end{align*}
\end{small}
Similar to $\bar{z}_i^\eta$, we take advantage of the anti-symmetric property of 
$Z_{u::}$ and assume $z_{uij} = - z_{uji}$ when $i > j$.

\section{Methods Considered in the Real World Application}
\label{sec:appendix:alternatives}

\begin{table*}[!h]
\caption{Specifics \& hyperparameters for models used when applying to real world datasets.}
\label{table:alternatives}
\begin{center}
\begin{small}
\begin{tabular}{lll}
    \toprule
    Model & Parameter & Values \\
    \midrule
    \multirow{5}{*}{SAD}  & Implementation & Supplementary Code\\
                          & Learning Rate  & $[0.001, 0.002, 0.005, 0.01, 0.02, 0.05, 0.1]$  \\
                          & \# Epochs      & $[2, 5, 10, 20, 50]$ \\
                          & $l_2$ Reg      & $[0.05, 0.01, 0.005, 0.001]$ \\
                          & $l_1$ Reg      & $0.01$\\
    \midrule
    \multirow{4}{*}{BPR \citep{rendle_bpr_2009}}  
                          & Implementation & Supplementary Code\\
                          & Learning Rate  & $[0.001, 0.002, 0.005, 0.01, 0.02, 0.05, 0.1]$  \\
                          & \# Epochs      & $[2, 5, 10, 20, 50]$ \\
                          & $l_2$ Reg      & $[0.05, 0.01, 0.005, 0.001]$ \\
    \midrule
    \multirow{4}{*}{SVD}  & Implementation & Surprise (\href{https://surprise.readthedocs.io/en/stable/matrix_factorization.html}{package}) \citep{Hug2020}\\
                          & Learning Rate  & $[0.001, 0.002, 0.005, 0.01, 0.02, 0.05, 0.1]$  \\
                          & \# Epochs      & $[2, 5, 10, 20, 50]$ \\
                          & Regularization & $[0.05, 0.01, 0.005, 0.001]$ \\
    \midrule
    \multirow{4}{*}{Matrix Factorization (MF)}
                          & Implementation & Cornac (\href{https://cornac.readthedocs.io/en/latest/index.html}{package}) \citep{salah2020cornac}\\
                          & Learning Rate  & $[0.001, 0.002, 0.005, 0.01, 0.02, 0.05, 0.1]$  \\
                          & \# Epochs      & $[2, 5, 10, 20, 50]$ \\
                          & $\lambda$ Reg  & $[0.05, 0.01, 0.005, 0.001]$ \\
    \midrule
    \multirow{4}{*}{\pbox{10cm}{Probabilistic Matrix Factorization (PMF) \\
                    \citep{mnih2008probabilistic}}} 
                          & Implementation & Cornac (\href{https://cornac.readthedocs.io/en/latest/index.html}{package})  \citep{salah2020cornac}\\
                          & Learning Rate  & $[0.001, 0.002, 0.005, 0.01, 0.02, 0.05, 0.1]$  \\
                          & \# Epochs      & $[2, 5, 10, 20, 50]$ \\
                          & $\lambda$ Reg  & $[0.05, 0.01, 0.005, 0.001]$ \\
    \midrule
    \multirow{4}{*}{Factorization Machine (FM) \citep{rendle2010factorization}} 
                          & Implementation & RankFM (\href{https://rankfm.readthedocs.io/en/latest/}{package}) \\
                          & Learning Rate  & $[0.001, 0.002, 0.005, 0.01, 0.02, 0.05, 0.1]$  \\
                          & \# Epochs      & $[2, 5, 10, 20, 50]$ \\
                          & $l_2$ Reg      & $[0.05, 0.01, 0.005, 0.001]$ \\
    \midrule
    \multirow{5}{*}{\pbox{10cm}{Neural Collaborative Filtering (NCF) \\ \citep{he2017neural}}}
                          & Implementation & MSFT recommenders (\href{https://github.com/microsoft/recommenders}{package}) \citep{graham2019microsoft} \\
                          & Learning Rate  & $[0.001, 0.002, 0.005, 0.01, 0.02, 0.05, 0.1]$  \\
                          & \# Epochs      & $[2, 5, 10, 20, 50]$ \\
                          & Batch size     & $[128, 256, 512, 1024]$  \\
                          & Network        & Three layers MLP with sizes $[128, 64, 32]$ \\
    \midrule
    \multirow{4}{*}{\pbox{10cm}{Variational AutoEncoder ($\beta$-VAE) \\ \citep{liang2018variational}}}
                          & Implementation & MSFT recommenders (\href{https://github.com/microsoft/recommenders}{package}) \citep{graham2019microsoft} \\
                          & $\beta$ parameter  & $[0.001, 0.002, 0.005, 0.01, 0.02, 0.05, 0.1]$  \\
                          & \# Epochs      & $[2, 5, 10, 20, 50]$ \\
                          & Batch size     & $[128, 256, 512, 1024]$  \\
    \bottomrule
\end{tabular}
\end{small}
\end{center}
\end{table*}

\section{Real World Examples}
\label{sec:appendix:examples}
During LOO evaluation, true preferences between a hold out item and the rest user interacted 
items in training set are determined based on users' actual ratings. Cases in which our 
model's predictions are consistent with true preferences while alternative models are not
are shown in Table \ref{table:appendix:examples}. We select $10$ such cases from each
of the three real world datasets. Model predictions together with item descriptions are 
shown in the table.

\newpage

\begin{landscape}
\begin{table*}[h]
  \caption{Examples where SAD produces a consistent prediction ($x_{uij} > 0$ \& $p_{uij} > 0.5$) while BPR fails ($x_{uij} < 0$ \& $p_{uij} < 0.5$).}
  \label{table:appendix:examples}
  \begin{small}
    \centering
    \begin{tabular}{llllcc}
    \toprule
    \multirow[b]{2}{*}{Dataset} & \multirow[b]{2}{*}{$u$-th user} & \multirow[b]{2}{*}{$i$-th item (rating)} & \multirow[b]{2}{*}{$j$-th item (rating)} & \multicolumn{2}{c}{$x_{uij} \mid p_{uij}$}  \\
    & & & & SAD & BPR \\
    \midrule
    \multirow[t]{10}{*}{Netflix} & '1381599' & Last of the Dogmen ($5$) & Look at Me ($2$) & $0.35 \mid 0.59$ & $-0.22 \mid 0.44$ \\
    \rowcolor{Gray} & '1243460' & Joe Kidd ($5$) & Scenes of the Crime ($1$) & $0.80 \mid 0.69$ & $-0.27 \mid 0.43$\\
    & '581011' & The Professional ($5$) & The Bourne Identity ($2$) & $0.80 \mid 0.69$ & $-0.32 \mid 0.42$ \\
    \rowcolor{Gray} & '632823' & Lara Croft: Tomb Raider: The Cradle of Life ($4$) & The Cookout ($2$) & $0.27 \mid 0.57$ & $-1.36 \mid 0.20$\\
    & '429299' & The Worst Witch ($4$) & A Chorus Line ($1$) & $0.85 \mid 0.70$ & $-0.76 \mid 0.32$ \\
    \rowcolor{Gray} & '127356' & Trading Spaces: Great Kitchen Designs and More! ($4$) & White Oleander ($2$) & $0.30 \mid 0.57$ & $-1.24 \mid 0.22$\\
    & '2264661' & Free Tibet ($5$) & Native American Medicine ($3$) & $1.25 \mid 0.78$ & $-0.05 \mid 0.49$ \\
    \rowcolor{Gray} & '581011' & The Dream Catcher ($4$) & The Bourne Identity ($2$) & $0.90 \mid 0.71$ & $-0.37 \mid 0.41$\\
    & '1368371' & Zombie Holocaust ($4$) & Gothika ($1$) & $0.71 \mid 0.67$ & $-0.31 \mid 0.42$ \\
    \rowcolor{Gray} & '1243460' & Moog ($3$) & Scenes of the Crime ($1$) & $0.98 \mid 0.73$ & $-1.34 \mid 0.21$\\
    \midrule
    \multirow[t]{10}{*}{Movie-Lens} & '1318' & High Fidelity ($5$) & Jimmy Hollywood ($3$) & $0.85 \mid 0.70$ & $-1.26 \mid 0.22$ \\
    \rowcolor{Gray} & '1250' & American Beauty ($5$) & eXistenZy ($3$) & $1.44 \mid 0.81$ & $-0.14 \mid 0.47$ \\
    & '4166' & My Fair Lady ($5$) & Problem Child 2 ($1$) & $0.51 \mid 0.63$ & $-0.78 \mid 0.31$ \\
    \rowcolor{Gray} & '153' & American Beauty ($4$) & Spice World  ($1$) & $1.07 \mid 0.75$ & $-0.08 \mid 0.48$ \\
    & '2160' & Harold and Maude ($4$) & The Brady Bunch Movie ($1$) & $0.34 \mid 0.58$ & $-0.73 \mid 0.32$ \\
    \rowcolor{Gray} & '4692' & Blade Runner ($5$) & The Newton Boys ($3$) & $0.65 \mid 0.66$ & $-0.38 \mid 0.41$ \\
    & '2385' & Braveheart ($4$) & Voyage to the Bottom of the Sea ($3$) & $0.98 \mid 0.73$ & $-0.56 \mid 0.36$ \\
    \rowcolor{Gray} & '4756' & Galaxy Quest ($3$) & Felicia's Journey ($2$) & $1.01 \mid 0.73$ & $-0.21 \mid 0.45$ \\
    & '4439' & The Maltese Falcon ($4$) & Entrapment ($3$) & $0.45 \mid 0.61$ & $-0.57 \mid 0.36$ \\
    \rowcolor{Gray} & '3021' & L.A. Confidential ($5$) & Fatal Attraction ($3$) & $0.64 \mid 0.66$ & $-0.37 \mid 0.41$ \\
    \midrule
    \multirow[t]{18}{*}{Food-Com} & '148323' & Best Ever Banana Cake \textbackslash w Cream Cheese Frosting ($5$) & Crock Pot Garlic Brown Sugar Chicken ($0$) & $0.16 \mid 0.54$ & $-3.51 \mid 0.03$ \\
    \rowcolor{Gray} & '424008' & Glazed Cinnamon Rolls, Bread Machine ($5$) & Japanese Mum's Chicken ($0$) & $0.31 \mid 0.57$ & $-0.38 \mid 0.41$ \\
    & '428423' & Crock Pot Stifado ($5$) & Best Ever and Most Versatile Muffins ($3$) & $0.30 \mid 0.57$ & $-2.99 \mid 0.05$ \\
    \rowcolor{Gray} & '733257' & Banana Banana Bread ($2$) & Low Fat Oatmeal Muffins ($0$) & $0.32 \mid 0.58$ & $-2.31 \mid 0.09$ \\
    & '340980' & Funky Chicken \textbackslash w Sesame Noodles ($3$) & Amanda's Thai Peanut ($1$) & $0.56 \mid 0.64$ & $-0.77 \mid 0.32$ \\
    \rowcolor{Gray} & '573772' & Delicious Chicken Pot Pie ($5$) & Amish Oven Fried Chicken ($1$) & $1.28 \mid 0.78$ & $-1.21 \mid 0.23$ \\
    & '1477540' & Cinnabon Cinnamon Rolls by Todd Wilbur ($4$) & Amanda's Cheese Pound Cake ($0$) & $0.81 \mid 0.69$ & $-2.59 \mid 0.07$ \\
    \rowcolor{Gray} & '268644' & Baked Tilapia \textbackslash w Lots of Spice ($5$) & Southern Fried Salmon Patties ($2$) & $0.38 \mid 0.59$ & $-3.35 \mid 0.03$ \\
    & '762742' & Easy Peezy Pizza Dough \& Bread Machine Pizza Doug ($5$) & Fresh Orange Muffins ($1$) & $0.98 \mid 0.73$ & $-2.31 \mid 0.09$ \\
    \rowcolor{Gray} & '212558' & Steak or Chicken Fajitas ($5$) & Thai Style Ground Beef ($3$) & $1.86 \mid 0.87$ & $-0.11 \mid 0.47$ \\
    \bottomrule
    \end{tabular}
  \end{small}
\end{table*}

\end{landscape}
%%%%%%%%%%%%%%%%%%%%%%%%%%%%%%%%%%%%%%%%%%%%%%%

\end{document}